\documentclass[times,tight]{aastex62}
\usepackage{amsfonts,times}
\usepackage{mathrsfs}
\usepackage{bm}
\usepackage{physics}
\usepackage{gensymb}
\usepackage{lipsum}
\usepackage[OT1]{fontenc} 

\def\cblue{}

\def\kms{\rm km\,s^{-1}}

\received{***}
\revised{***}
\accepted{***}

\shorttitle{Quasar distances measurements by SARM}
\shortauthors{Songsheng et al.}

\begin{document}

\title{\large \bf Geometric Distances of Quasars Measured by 
Spectroastrometry and Reverberation Mapping:  Monte Carlo Simulations}
\correspondingauthor{Jian-Min Wang}
\email{wangjm@ihep.ac.cn}

\author{Yu-Yang Songsheng}
\affil{Key Laboratory for Particle Astrophysics,
Institute of High Energy Physics,
Chinese Academy of Sciences,
19B Yuquan Road, Beijing 100049, China}
\affil{University of Chinese Academy of Sciences,
19A Yuquan Road, Beijing 100049, China}

\author{Yan-Rong Li}
\affil{Key Laboratory for Particle Astrophysics,
Institute of High Energy Physics,
Chinese Academy of Sciences,
19B Yuquan Road, Beijing 100049, China}

\author{Pu Du}
\affil{Key Laboratory for Particle Astrophysics,
Institute of High Energy Physics,
Chinese Academy of Sciences,
19B Yuquan Road, Beijing 100049, China}

\author{Jian-Min Wang}
\affil{Key Laboratory for Particle Astrophysics,
Institute of High Energy Physics,
Chinese Academy of Sciences,
19B Yuquan Road, Beijing 100049, China}
\affil{University of Chinese Academy of Sciences,
19A Yuquan Road, Beijing 100049, China}
\affil{National Astronomical Observatories of China,
Chinese Academy of Sciences,
20A Datun Road, Beijing 100020, China}

\begin{abstract}
Recently, GRAVITY onboard the Very Large Telescope Interferometer (VLTI) first 
spatially resolved the structure of the quasar 3C 273 with an unprecedented 
resolution of $\sim 10\mu$as. A new method of measuring parallax distances has 
been successfully applied to the quasar through joint analysis of 
spectroastrometry (SA) and reverberation mapping (RM) observation of its broad 
line region (BLR). The uncertainty of this SA and RM (SARM) measurement is 
about $16\%$ from real data, showing its great potential as a powerful tool for 
precision cosmology. In this paper, we carry out detailed analyses of mock data 
to study impacts of data qualities of SA observations on distance measurements 
and establish a quantitative relationship between statistical uncertainties of 
distances and relative errors of differential phases. We employ a 
circular disk model of BLR for the SARM analysis. We show that SARM analyses of 
observations generally generate reliable quasar distances, even for relatively 
poor SA measurements with error bars of $40\%$ at peaks of phases. Inclinations 
and opening angles of BLRs are the major parameters governing distance 
uncertainties. It is found that BLRs with inclinations $\gtrsim 10\degr$ and 
opening angles $\lesssim 40\degr$ are the most reliable regimes from SARM 
analysis for distance measurements. Through analysis of a mock sample 
of AGNs generated by quasar luminosity functions, we find that if the GRAVITY/
GRAVITY+ can achieve a phase error of $0.1\degr$ per baseline for targets with 
magnitudes $K \lesssim 11.5$, the SARM campaign can constrain $H_0$ 
to an uncertainty of $2\%$ by observing $60$ targets.
\end{abstract}


\section{Introduction}
The distance of a celestial object can be in principle measured through the 
geometric relation $D_{\rm A}=\Delta R/\Delta \theta$, where $\Delta R$ and 
$\Delta \theta$ are its linear and angular size, respectively. However, it is 
extremely hard to measure both $\Delta R$ and $\Delta\theta$ for the same 
object, in particular those at cosmological distances. Either $\Delta R$ is too 
large or $\Delta\theta$ is too small. At cosmological distances, only active 
galactic nuclei (AGNs) and quasars can be feasibly measured for both $\Delta R$ 
and $\Delta\theta$ of their broad-line regions (BLRs) owing to the breakthrough 
progress of high spatial resolution and reverberation mapping (RM) of AGNs and 
quasars nowadays, respectively. Thanks are given to GRAVITY, an interferometric 
instrument operating in the $K$-band at the Very Large Telescope Interferometry 
(VLTI), for its unprecedented high spatial resolution through spectroastrometry 
(SA) \citep{eisenhauer2008,gravity2017}, making it feasible to measure angular 
sizes of BLRs of AGNs. Recently, a direct measurement of angular diameter of 
BLR of quasar 3C 273 through SA of VLTI \citep{gravity2018} reaches a spatial 
resolution of $\sim10\mu$as, successfully revealing a flattened and Keplerian 
rotating disk-like structure of the BLR. Along with the measurement of its 
linear size by a long-term RM campaign of 3C 273 \citep{zhang2019}, \cite
{wang2020} made a joint analysis of SA and RM (SARM) data and obtained the 
first parallax distance for the quasar with $16\%$ precision. This is a 
compelling effort for quasar distances and shines a light on a new way for 
cosmology of the Hubble constant, which arises an intensive debate currently 
known as $H_{0}$-tension \citep{riess2019}.

Broad emission lines with full-width-half-maximums (FWHMs) ranging from $10^3$ 
to $10^4{\rm\,km\,s^{-1}}$ are the prominent features of type I AGN and quasar 
spectra. They are from fast moving clouds in BLRs photoionized by ionizing 
radiation from accretion disks around central supermassive black holes (SMBHs) 
\citep{lyndenbell1969,rees1984}. Variation in the strength and profile of the 
broad line will follow the ionizing continuum, but with a delay because of 
different paths of the broad line and ionizing photons. The delay is 
approximately equal to the light travel time from the central source to the 
BLR. This is known as the RM of the BLR in AGN \citep{blandford1982}.
{\cblue Physical sizes of BLRs can be simply estimated by cross-correlation 
functions (CCFs) between light curves of continuum and broad emission lines 
\citep{peterson1993}. RM campaigns spectroscopically monitoring AGNs can futher 
probe geometries and dynamics of BLRs as well as SMBH masses by more advanced 
techniques such as velocity resolved CCF \citep{bentz2010}, transfer function 
recovery \citep{horne1994} and dynamical modeling \citep{pancoast2011}.}
Over the past few decades, about $120\sim 150$ AGNs with high quality data have 
been measured for their sizes and black hole mass through AGN Watch \citep
{peterson1998,bentz2013}, Bok2.4 \citep{kaspi2000}, SEAMBHs (Super-Eddington 
Accreting Massive Black Holes) \citep{du2019,hu2021},  MAHA (Monitoring AGNs 
with H$\beta$ Asymmetry) \citep{du2018,brotherton2020}, and SDSS-RM projects 
\citep{shen2016,grier2017,fonseca2020}. More targets are being monitored and 
expected to be reported in next few years.

For most AGNs, sizes of their BLRs range from a few light days to a few hundred 
light days, and their angular diameters hardly exceed $\sim100 {\rm\,\mu as}$, 
well below the imaging resolution of currently available facilities. 
Fortunately, due to the bulk motion (e.g., rotation, inflow or outflow) of the 
BLR gas, the gas moving toward observers and the gas moving away from observers 
are distributed in different spatial positions, making it possible to apply the 
super-resolution capability of SA to BLRs \citep{petrov2001,marconi2003}. For a 
source whose global angular size is smaller than the interferometer resolution, 
the interferometric phase is proportional to its angular displacements along 
the projected baseline from the direction where the optical path difference to 
the pair of telescopes vanishes \citep{petrov1989}. SA measures the 
interferometric phase as a function of wavelength near the broad emission line, 
thus obtains angular displacements of clouds with different line-of-sight (LOS) 
velocities. The angular size, geometry and dynamics of BLR can be probed in 
this way.

Through a joint analysis of SARM data, the linear and angular size of the BLR 
can be determined simultaneously, providing a parallax measurement of AGN's 
distance. Such a measurement does not need calibration of any cosmic distance 
ladder or correction of extinction and reddening, providing a geometric way to 
measure the Hubble constant. The first SARM analysis of a single object 3C 273 
obtains $H_0 = 71.5^{+11.9}_{-10.6} {\rm\,km\,s^{-1}\,Mpc^{-1}}$ with precision 
$16\%$ \citep{wang2020}. Moreover, based on the current capabilities of 
GRAVITY, about $50$ AGNs are expected to be observed through SA, bringing down 
the uncertainty of $H_0$ below $2.5\%$ \citep{wang2020}. Future SA observations 
of fainter AGNs across a wider redshift range through powerful facilities 
(GRAVITY+/VLTI, and Extremely Large Telescope) will be feasible. Not only can 
the Hubble constant be determined more precisely, but also a wider 
distance-redshift relation can be established to look back the expansion 
history of our Universe. Cosmology will be greatly advanced by this effort.

As a natural step to approach quasar distances through SARM observations, it is 
necessary to simulate mock data using Monte Carlo method and quantify 
uncertainties of distance measurements under varied data qualities and object 
properties. It will be a guidance for arranging the observation campaign and 
selecting appropriate targets to minimize uncertainties. This paper is 
structured as follows. Section 2 presents the framework used to generate mock 
data and estimate the input model parameters. Section 3 analyzes and summarizes 
simulation results. Discussions are provided in section 4, and conclusions in 
the last section.

\section{Methodology}
In order to study impacts of various factors on the constraint ability of SARM 
campaign on cosmology, we will use a parameterized BLR model to generate mock 
data of SARM observations. Appendix \ref{sec:mock data} presents details of 
mock data generations. Reconstruction of the BLR model can be done through mock 
data fitting via the diffusive nested sampling (DNest) algorithm
\footnote{
    The DNest algorithm was proposed by \cite{brewer2011} and an implementation 
    package developed by the authors is available at \url{https://github.com/
    eggpantbren/DNest4}. In this work, we use our own DNest package \texttt
    {CDNest} \citep{li2020} that is written in C language and enables the 
    standarized parallel message passing interface, which is available at \url
    {https://github.com/LiyrAstroph/CDNest}.}
\citep{brewer2018}, obtaining the probability distribution of model 
parameters, including the angular distance of the AGN. It is our goal to 
analyze how uncertainties and degeneracies of reconstructed parameters change 
with relative errors of data and values of input parameters. The result can be 
a guidance for object selection and observation strategy in future SARM 
campaign to meet the need of cosmology.
{\cblue The method to generate and fit SARM data is implemented in \texttt{BRAINS}\footnote{\url{https://github.com/LiyrAstroph/BRAINS}}.}

\subsection{Parameterized BLR model}\label{sec:model}
\begin{figure}
    \plotone{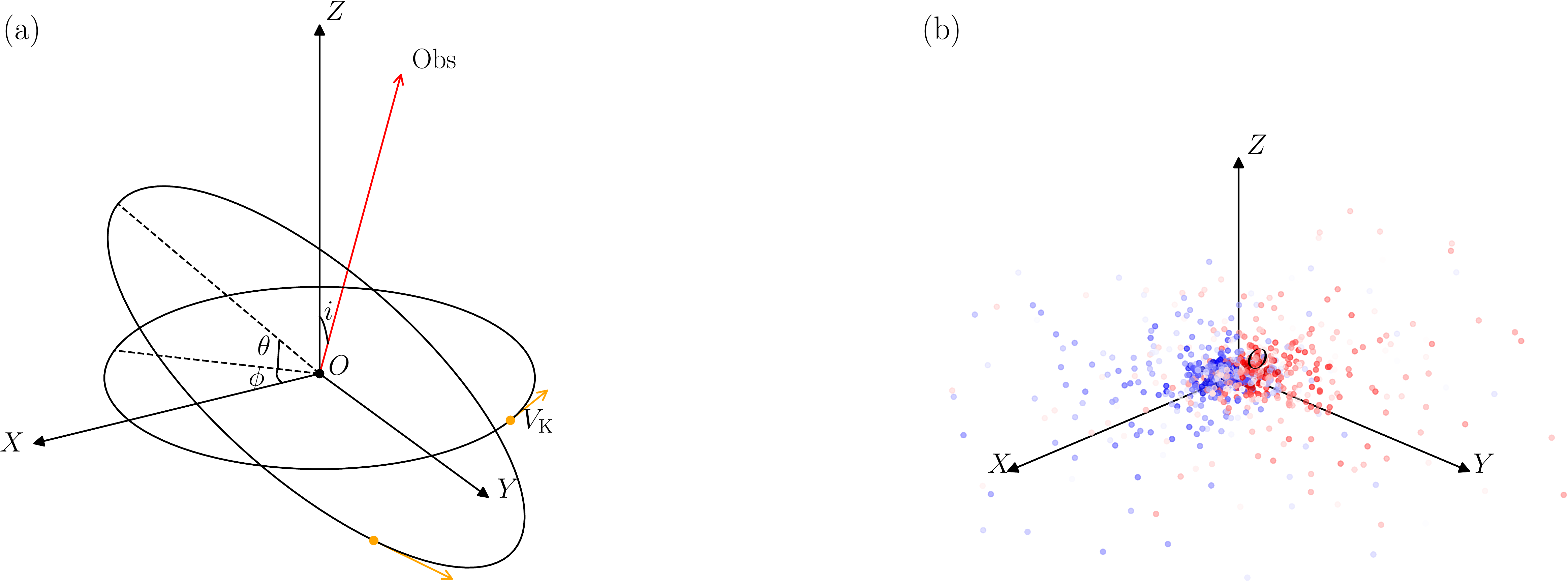}
    \caption{Illustration of the BLR model.
    (a) Coordinate system used in the present model. $O$ is the central black 
    hole and $O-XY$ is equatorial plane of the BLR. A remote observer with an 
    inclination of $i$ to the $OZ$-axis is located in the $O-YZ$ plane. All 
    clouds are on circular orbits around the central black hole with Keplerian 
    velocities $V_{\rm K}$. {\cblue Two ellipses in the fugure show two 
    possible orbits, and the orange arrow represents the velocity of a cloud at 
    that position.} $\theta$ is the angle between the orbital plane and 
    the equatorial plane, and $\cos\theta$ is uniformly distributed between $
    [\cos\theta_{\rm opn},1]$, where $\theta_{\rm opn}$ is the half opening 
    angle of the BLR. $\phi$ is the azimuthal angle of the orbital plane and 
    uniformly distribute between $[0,2\pi]$.
    (b). Distribution of clouds generated for a typical BLR. Colors of clouds represent line of sight velocities. {\cblue The observer's line of sight is perpendicular to the paper. Clouds with red color are moving away from the observer while those with blue color towards the observer.}
    There is an obvious velocity gradient in space, which is crucial for SA observations.
    \label{fig:coord}}
\end{figure}

Over the past few decades, a sample of more than 100 AGNs with RM observations 
have advanced our understanding of BLRs a lot \citep{peterson1998,kaspi2000, 
bentz2013,du2018}. Particularly, geometries and dynamics of BLRs can be well 
studied through velocity-resolved delay maps obtained by RM \citep{grier2013}. 
A disk-like BLR is common in many broad line Seyfert 1 galaxies \citep 
{bentz2013,grier2013,lu2016,du2016,xiao2018}, even in some narrow-line Seyfert 
1 galaxies \citep{du2016}. Multiple RM observations of several objects, such as 
NGC 5548, 3C 390.3, and NGC 7469, show that the FWHM of H$\beta$ line and its 
lag $\tau_{\rm H\beta}$ follows $\tau_{\rm H\beta} \propto {\rm FWHM}^{-1/2}$, 
indicating Keplerian rotation of the BLR gas \citep{peterson2004}. A flattened 
disk-like BLR in 3C 273 has also been detected by GRAVITY. Disk-like BLR with 
Keplerian rotation could be common.

Parameterized BLR model has been applied widely in RM data fitting in order to 
obtain physical quantities, such as BLR size and black hole mass, in a 
self-consistent way \citep{pancoast2014b,li2018,williams2018}. A comprehensive 
model with $\sim 30$ parameters has been developed to include complex 
geometries and dynamics for perfect fitting of the observation data \citep
{pancoast2014a}. But as a preliminary guidance for SARM, we reasonably assume 
the Keplerian rotating disk model for the BLR, and include only the fundamental 
parameters for concise estimations of parameter impacts. This model also works 
quite well in \cite{gravity2018} and \cite{wang2020} when fitting SA and SARM 
data of 3C 273, respectively.

The BLR is composed of a large amount of line-emitting clouds on circular 
orbits around the central black hole with Keplerian velocities, as shown in 
Fig. \ref{fig:coord}. The radial distribution of clouds is described by a 
shifted $\Gamma$-distribution. The distance of BLR clouds to the SMBH is 
generated by
\begin{equation}\label{eq:gamma1}
    r = R_{\rm S} + \mathscr{F} R_{\rm BLR} + \Gamma_0 \beta^2 (1 - \mathscr{F}) R_{\rm BLR},
\end{equation}
where $R_{\rm S}$ is the Schwarzschild radius, $R_{\rm BLR}$ is the mean 
radius, $\mathscr{F} = R_{\rm in} / R_{\rm BLR}$ is the fraction of the inner 
to the mean radius, $\beta$ is the shape parameter, and $\Gamma_0 = p(x|\beta^
{-2},1)$ is a random number drawn from a $\Gamma$-distribution
\begin{equation}\label{eq:gamma2}
    p(x|\alpha,x_0) = \frac{x^{\alpha-1}\exp(-x/x_0)}{x_0^{\alpha}\Gamma(\alpha)},
\end{equation}
where $x_0$ is a scale factor ($x_0 = 1$ here), $\alpha = \beta^{-2}$ and  
$\Gamma(\alpha)$ is the $\Gamma$-function. Symbols and meanings of all free 
parameters used in the BLR model are summarized in Table \ref{tab:BLR}.

Our model in this work differs a little with that in \cite{gravity2018} and 
\cite{wang2020} in the orientation distribution of orbital planes. We assume 
that the cosine of the angle between the direction of orbital angular momentum 
and $Z$-axis is uniformly distributed over $[\cos\theta_{\rm opn}, 1]$ to reach 
a uniform distribution of clouds. In previous work, the angle between the 
direction of angular momentum and $Z$-axis is uniformly distributed over $[0, 
\theta_{\rm opn}]$, and so clouds will be accumulated near the equatorial plane.
We make the change for a more well-defined definition of opening angles. 
{\cblue Clouds are randomly distributed along a given orbit by assigning their 
orbital phases uniformly over $[0, 2\pi]$.}
\begin{table*}
    \footnotesize
    \centering
    \caption{\label{tab:BLR} Parameters used in the BLR model}
    \begin{tabular}{lllll}\hline\hline
        Parameters & Meanings & Prior ranges & Fiducial values & Simulation ranges \\ \hline
        $D_{\rm A} {\rm (Mpc)}$ & Angular distance & $[10,10^4]$ & $42.555$ &  \\
        $i (\degr)$ & Inclination angle of the LOS & $[0,90]$ & $25$ & $[1,45]$\\
        $\rm PA (\degr)$ & Position angle & $[0, 360]$ & $90$ & $[0, 360]$\\
        $M_{\bullet} (M_{\sun})$ & Supermassive black hole mass & $[10^6, 10^9]$ & $2\times10^7$ &  \\
        $R_{\rm BLR} {\rm (ld)}$ & Mean linear size & $[1,10^3]$ & $15$ & \\
        $F$ & Fractional inner radius & $[0,1]$ & $0.25$ & $[0.1,0.9]$\\
        $\beta$ & Radial distribution shape parameter & $[0,4]$ & $1.5$ & $[0.5,2.5]$ \\
        $\theta_{\rm opn} (\degr)$ & Half opening angle & $[0,90]$ & $25$ & $[1,45]$\\
    \hline    
    \end{tabular}
\end{table*}

\subsection{Spectroastrometry}
A detailed mathematical formulation of SA of BLR can be found in \cite
{rakshit2015} and \cite{songsheng2019b}. We summarize it here for reader's 
convenience. For an interferometer with a baseline $\bm{B}$, the differential 
interferometric phase for a non-resolved source is
\begin{equation}\label{eq:phase}
    \phi_{*}(\lambda) = -2\pi \bm{u} \vdot [\bm{\epsilon}(\lambda) - \bm{\epsilon}(\lambda_{\rm r})],
\end{equation}
where $\bm{u}=\bm{B}/\lambda$ is the spatial frequency, $\bm{\epsilon}$ is the 
photocenter of the source at wavelength $\lambda$, $\lambda_{\rm r}$ is the 
wavelength of a reference channel. Here the bold letters represents vectors. 
Given the surface brightness distribution $\mathcal{O}(\bm{\alpha},\lambda)$ of 
the source, we have
\begin{equation}\label{eq:photocenter}
    \bm{\epsilon}(\lambda) = \frac{\int \bm{\alpha} \mathcal{O}(\bm{\alpha},\lambda)  \dd[2]{\bm{\alpha}}}{\int \mathcal{O}(\bm{\alpha},\lambda) \dd[2]{\bm{\alpha}}},
\end{equation}
where $\bm{\alpha}$ is the angular displacement on the celestial sphere. For 
AGNs, $\mathcal{O} = \mathcal{O}_{\ell} + \mathcal{O}_{\rm c}$, where $\mathcal
{O}_{\ell}$ and $\mathcal{O}_{\rm c}$ are the surface brightness distribution 
contributed by the BLR and continuum regions, respectively. Once the BLR model 
is set up, $\mathcal{O}_{\ell}$ can be calculated through
\begin{equation}\label{eq:brightness}
    \mathcal{O}_{\ell}(\bm{\alpha},\lambda) = \int \frac{\Xi_r F_{\rm c}}{4 \pi 
    r^2} f(\bm{r},\bm{V}) \var(\bm{\alpha}-\bm{\alpha}') \var(\lambda -\lambda^
    {\prime}) \dd[3]{\bm{r}} \dd[3]{\bm{V}},
\end{equation}
where $\lambda^{\prime}=\lambda_{\rm cen} \gamma_{0}\left(1+\bm{V}\vdot\bm{n}_ 
{\rm obs}/c\right) \left(1-R_{\rm S}/r\right)^{-1/2}$ is the shifted wavelength 
of the photon from the broad emission line centered at $\lambda_{\rm cen}$ by 
Doppler effect and gravitation, $\gamma_{0}=\left(1-V^2/c^2\right)^{-1/2}$ is 
the Lorentz factor, ${\bm \alpha}^{\prime} = \left[\bm{r} - \left(\bm{r}\vdot\bm
{n}_{\rm obs}\right) \bm{n}_{\rm obs}\right] / D_{\rm A}$, $\bm{r}$ is the 
displacement to the central BH, $\Xi_r$ and $f(\bm{r},\bm{V})$ are the 
reprocessing coefficient and velocity distribution of the clouds at position 
$\bm{r}$ respectively, $F_{\rm c}$ is ionizing fluxes received by an observer, 
$\bm{n}_{\rm obs}$ is the unit vector pointing from the observer to the source 
and $D_{\rm A}$ is the angular size distance of the AGN. Introducing the 
fraction of the emission line to total ($\ell_{\lambda}$), we have
\begin{equation}\label{eq:fraction}
    \bm{\epsilon}(\lambda) = \ell_{\lambda}\,\bm{\epsilon}_{\ell}(\lambda),
\end{equation}
where
\begin{equation}
    \bm{\epsilon}_{\ell}(\lambda) = \frac{\int \bm{r} \mathcal{O}_{\ell} \dd[2]
    {\bm{\alpha}}}{\int \mathcal{O}_{\ell} \dd[2]{\bm{\alpha}}} \qc
    \ell_{\lambda} = \frac{F_{\ell}(\lambda)}{F_{\rm tot}(\lambda)} \qc
    F_{\ell}(\lambda) = \int \mathcal{Q}_{\ell} \dd[2]{\bm{\alpha}} \qc
    F_{\rm tot}(\lambda)=F_{\ell}(\lambda) + F_{\rm c}(\lambda). \nonumber
\end{equation}
Differential phase curve (DPC) can be obtain by inserting Eq. (\ref{eq:fraction}), (\ref{eq:brightness}) and (\ref{eq:photocenter}) into (\ref{eq:phase}). If 
the DPC whose amplitudes are a few degrees can be measured with SA, the 
effective spatial resolution will be 100 times the resolution limit $\lambda / 
B$.

\subsection{Reverberation mapping} 
A detailed mathematical formulation of one dimensional RM can be found in \cite 
{li2013,li2018}. We also summarize it here for reader's convenience. Damped 
random walk (DRW) model is used to describe continuum variations in order to 
interpolate and extrapolate light curves of continuum \citep{kelly2009,zu2013}. 
We first express the continuum light curve $\bm{y}$ by $\bm{y} = \bm{s} + \bm 
{n} + \bm{E}q$, where $\bm{s}$ denotes the underlying signal of the variation, 
$\bm{n}$ represents the measurement noise, $q$ is the mean value of the light 
curve, and $\bm{E}$ is a vector whose elements are all $1$. In the DRW model, 
the covariance function between $s_i$ and $s_j$ is given by 
\begin{equation}
    S(t_i,t_j) = \sigma_{\rm d}^2 \exp(-\frac{|t_i-t_j|}{\tau_{\rm d}}), 
\end{equation} 
where $t_i$ and $t_j$ are time for signal $s_i$ and $s_j$, respectively, 
$\sigma_{\rm d}$ is the long-term standard deviation of the variation and $\tau_
{\rm d}$ the typical timescale.

We further assume that both $\bm{s}$ and $\bm{n}$ are Gaussian and 
uncorrelated. Given $\bm{y}$, posterior distribution of $\sigma_{\rm d}$, $\tau_
{\rm d}$ and $q$ can be obtained by Bayesian analysis with likelihood function
\begin{eqnarray}
    P(\bm{y}|\sigma_{\rm d},\tau_{\rm d},q) = \frac{1}{\sqrt{(2\pi)^m |\bm{C}|}} \exp\left[-\frac{(\bm{y}-\bm{E}q)^T \bm{C}^{-1} (\bm{y}-\bm{E}q)}{2}\right],
\end{eqnarray}
where superscript ``$T$'' denotes transposition, $\bm{C} = \bm{S} + \bm{N}$, 
$\bm{S}$ and $\bm{N}$ are the covariance matrix of $\bm{s}$ and $\bm{n}$ 
respectively, and $m$ is the number of data points.

Given parameters $(\sigma_{\rm d},\tau_{\rm d},q)$, a typical realization for 
the observed continuum light curve will be
\begin{equation}
    f_c = (\bm{u}_s + \hat{\bm{s}}) + \bm{E}(u_q - \hat{q}), 
\end{equation}
where $\hat{q} = \bm{E}^T \bm{C}^{-1} \bm{y} / \bm{E}^T \bm{C}^{-1} \bm{E}$, 
$\hat{\bm{s}} = \bm{S}^T \bm{C}^{-1} (y - \bm{E}\hat{q})$, and $\bm{u}_s$ and 
$u_q$ are Gaussian processes with zero mean and covariance matrices $\bm{Q} = 
(\bm{S}^{-1} + \bm{N}^{-1})^{-1}$ and $C_q = (\bm{E}^T\bm{C}^{-1}\bm{E})^{-1}$ 
respectively. Next, $\bm{u}_s$ and $u_q$ are treated as free parameters and 
further constrained by light curves of the emission line.

Given the BLR model and the realization of continuum light curves, the  
variation of the emission line is calculated by
\begin{equation}
    f_{\ell}(t) = \int \dd{\bm{r}} \dd{t^{\prime}} \frac{\Xi_{r}f_{\rm c}(t^{\prime})}{4\pi r^{2}} n(\bm{r}) 
    \delta\left(t^{\prime}-t+\tau\right),
\end{equation}
where $\tau = (r - \bm{r}\vdot\bm{n}_{\rm obs})/c$, $\Xi_{r}$ is the 
reprocessing coefficient and $\bm{n}(\bm{r})$ is the number density of the 
clouds.

\subsection{Joint analysis}\label{sec:joint analysis}
The goal of a fully joint analysis is to obtain the posterior probability 
distribution of the model parameters consistently using the combined data from 
SA and RM observations. 
{\cblue SA and RM observations are conducted independently, thus we assume} that the probability distribution for the measurement errors of light curves, profiles and DPCs are uncorrelated. The joint likelihood function can be written as
\begin{eqnarray}
    P(\mathscr{D} | \bm{\Theta}) 
    &=\displaystyle\prod_{i=1}^{N_{\rm RM}} \frac{1}{\sqrt{2 \pi\sigma_{\ell}^{2}}} 
    \exp \left\{-\frac{\left[f_{\rm \ell,obs}-f_{\rm\ell,mod}\left(f_{\rm c, obs} | \bm{\Theta}\right)\right]^{2}}{2\sigma_{\ell}^{2}}\right\}
    \times \prod_{i=1}^{N_{\rm G}} \prod_{j=1}^{N_{\lambda}}  \frac{1}{\sqrt{2 \pi \sigma_{\phi_{ij}}^{2}}} 
    \exp \left\{-\frac{\left[\phi_{\rm obs}-\phi_{\rm mod} \left(\bm{\Theta}\right)\right]^{2}}{2\sigma_{\phi_{ij}}^{2}}\right\} \nonumber \\
    &\displaystyle\times \prod_{j=1}^{N_{\lambda}} \frac{1}{\sqrt{2 \pi \sigma_{\rm F}^{2}}} 
    \exp \left\{- \frac{\left[F_{\rm \ell,obs} - F_{\rm\ell,mod}\left(\bm{\Theta}\right)\right]^{2}}{2\sigma_{\rm F}^{2}}\right\},
\end{eqnarray}
where $\mathscr{D}$ represents the data set, $\bm{\Theta}$ represents all the 
model parameters, $f_{\rm c, obs}$ is the continuum data, $f_{\rm \ell,obs}$, 
$F_{\rm\ell,obs}$, and $\phi_{\rm obs}$ are the line flux, line profile, and 
DPC of the emission line with measurement uncertainties $\sigma_{\ell}$, 
$\sigma_{\phi_{ij}}$ and $\sigma_{\rm F}$ respectively, and $(f_{\rm\ell,mod},F_
{\rm\ell,mod},\phi_{\rm mod})$ are the corresponding predicted values from the 
BLR model.

In light of Bayes' theorem, the posterior probability distribution for $\bm
{\Theta}$ is given by 
\begin{equation}\label{eq:posterior} 
P(\bm{\Theta} | \mathscr{D}) = \frac{P(\bm{\Theta}) P(\mathscr{D} | \bm{\Theta})}{P(\mathscr{D})},
\end{equation}
where $P(\bm{\Theta})$ is the prior distribution of the model parameter and $P(\mathscr{D})$ is a normalization factor. DNest algorithm can be applied to sample the distribution Eq. (\ref{eq:posterior}).
{\cblue A brief introduction to the DNest algorithm is given in Appendix \ref{sec:DNest}.}

\section{Results}
\subsection{Impacts of data qualities}\label{sec:qualities}
\begin{figure}
    \gridline{\fig{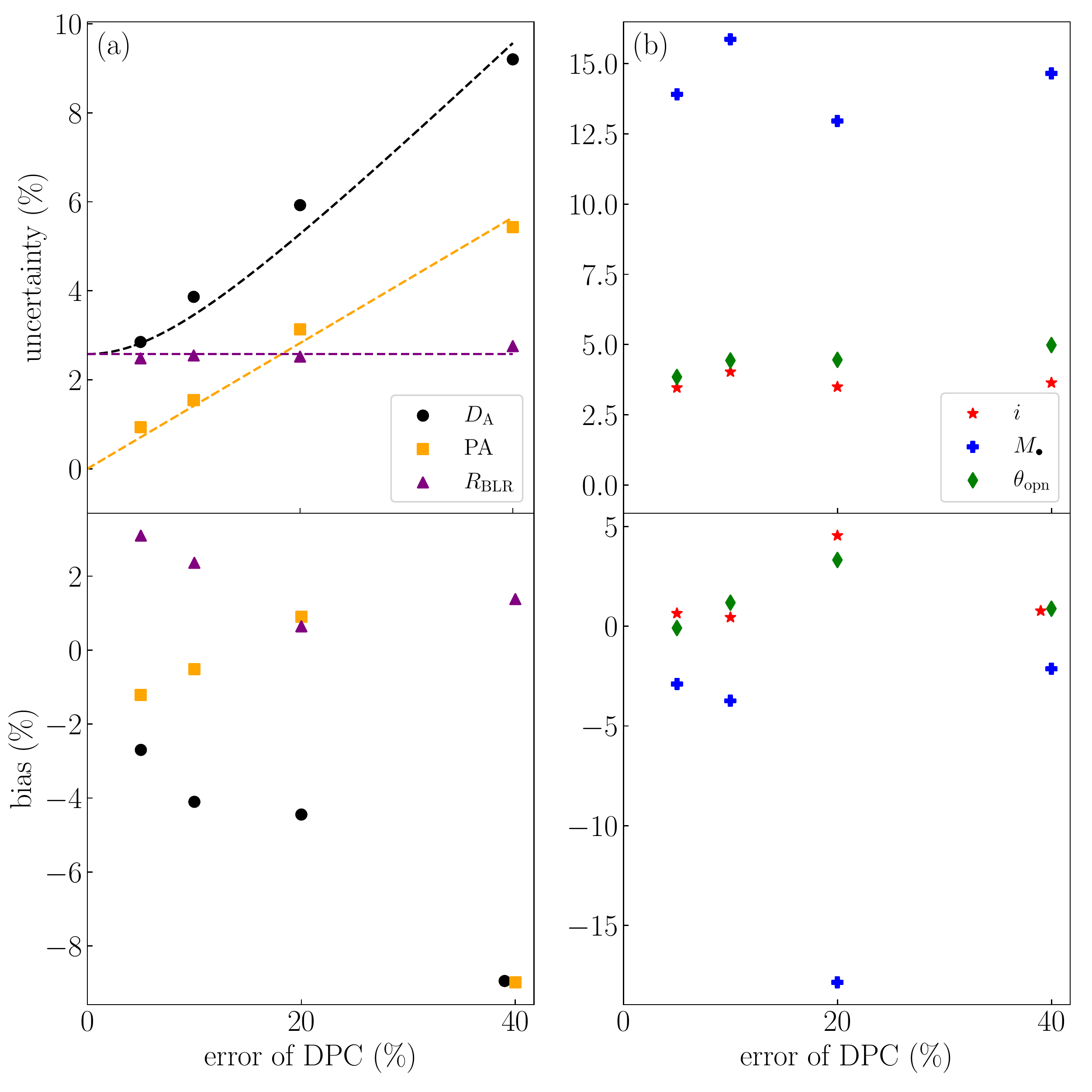}{0.7\textwidth}{}}
    \caption{Relative uncertainties and bias of part of model parameters under 
    different SA data qualities. The mock data of SA and RM are both generated 
    by the BLR model of the same parameter values shown in Table \ref{tab:BLR}. 
    Relative errors of RM data are fixed, while those of SA data are varied to 
    be $5\%$, $10\%$, $20\%$ and $40\%$. The configuration of baselines follows 
    the SA observation of 3C 273. So the target is observed four times for SA 
    data. \label{fig:sim_error}}
\end{figure}

SARM data consists of light curves of optical continuum and emission lines, 
profiles of NIR emission lines as well as its DPC. Observation of light curves 
and line profiles are relatively mature and data quality for most SARM targets 
can be quiet good. On the contrary, techniques for obtaining DPCs are still in 
development, and data quality largely depends on seeing and target brightness. 
In this subsection, we mainly study the impact of DPC data quality on the 
measurement of distance, black hole mass and BLR geometry.

To generate typical mock data, we take the fiducial values in Table \ref
{tab:BLR} for the BLR model. The light curves of continuum and emission line 
last for 200 days with 1 day cadence. The continuum light curves are generated 
by the DRW model with $\tau_{\rm d} = 60 {\rm\,d}$ and $\sigma_{\rm d} = 0.25$. 
The relative errors of continuum and line variations are $0.5\%$ and $1\%$, 
respectively. 
{\cblue We generate mock RM data with higher qualities than those in typical RM campaigns in order to focus on impacts of data qualities of SA observations on distance measurements.}
The redshift of the target is assumed to be $0.01$, and the NIR 
emission lines for SA is Brackett $\gamma$ (Br$\gamma$) centered at $2.166 
{\rm\,\mu m}$ in rest frame. The equivalent width of Br$\gamma$ is $40$ \AA. We 
follow the spectral resolution of GRAVITY ($\lambda/\Delta\lambda\sim 500$) and 
broaden the profile by a Gaussian with ${\rm FWHM} = 4{\rm\,nm}$ ($\sim 600 
\kms$). The relative error of the profile is $0.5\%$. The configuration of 
baseline we used for generating DPC is the same as that of Extended Data Fig. 1
(b) in \citet{gravity2018} (which means the target is observed four times in 
total). We also assume the absolute error of phase isthe same in all wavelength 
channels and baselines for simplification.

To quantify data qualities of DPCs, we define the relative error as the ratio 
of phase error to the peak of DPCs with largest amplitudes. We vary the 
relative error of DPC to be $5\%$, $10\%$, $20\%$ and $40\%$. For each data 
set, we obtain the posterior probability distribution of model parameters 
through Bayesian analysis introduced in subsection \ref{sec:joint analysis}. 
An example of generation and fitting of mock data is illustrated in Appendix \ref{sec:mock data}. 

We define inferred value of the model parameters as median of the posterior 
distribution, and lower/upper bound as $16\%$ and $84\%$ quantile. Uncertainty 
of the parameters are defined as half of the difference between the upper and 
lower bounds and bias as the difference between the inferred and input value. 
Finally, for dimensional quantities, we divide the uncertainty by inferred 
value and bias by input value to get the relative uncertainty and bias; for 
angles, we divide them by $1 {\rm rad}$ instead. We emphasis here that bias is 
not systematic uncertainty. It is the deviation of inferred values of model 
parameters from input ones caused by finite width of posterior probability 
distribution (uncertainty). The absolute value of bias is comparable to 
uncertainty usually. A much larger bias than uncertainty indicates the fail of 
Bayesian analysis, mostly caused by degeneracy between model parameters.

Relative uncertainties and bias of part of model parameters under different SA 
data qualities are shown in Fig. \ref{fig:sim_error}. As we can see in upper 
panel of Fig. \ref{fig:sim_error}(a), relative uncertainties of $R_{\rm BLR}$ 
are $2.58\%$ and independent of errors of DPCs, since it is only constrained by 
light curves of continuum and emission lines. Uncertainties of $\rm PA$ are 
proportional to errors of DPCs and the proportional coefficient is 
approximately $0.14$. Since $D_{\rm A} = R_{\rm BLR} / \xi_{\rm BLR}$, where 
$\xi_{\rm BLR}$ is the mean angular size, we have
\begin{equation}
    \delta_{D_{\rm A}} = \sqrt{\delta_{R_{\rm BLR}}^2 + \delta_{\xi_{\rm BLR}}^2},
\end{equation}
where $\delta_x$ means the relative uncertainty of $x$. If we assume $\delta_
{\xi_{\rm BLR}} = k\delta_{\rm DPC}$ and let $\delta_{R_{\rm BLR}} = 
2.58\%$, we can get through fitting
\begin{equation}
    \delta_{D_{\rm A}} = \sqrt{(2.58\%)^2 + (0.23\delta_{\rm DPC})^2}.
\end{equation}
We must emphasis that this relation is valid only for this set of parameters, 
since relative uncertainties of $D_{\rm A}$ also depends on inclinations and 
opening angles, as shown in the next subsection. There are no systematic 
uncertainties for mock data analysis, but they should be considered in
real observations.

Lower panel of Fig. \ref{fig:sim_error}(a) presents the relation between bias 
and data error. Median values of posterior distributions are obtained with 
greater randomness than uncertainties, leading to a more irregular pattern. The 
bias of $R_{\rm BLR}$ changes little with data error, while the bias of $D_{\rm 
A}$ increases with data error, in agreement with results in the upper panel. 
The bias of $\rm PA$ fluctuates when $\delta_{\rm DPC} \le 20\%$, but increases 
a lot when $\delta_{\rm DPC} = 40\%$.
\begin{figure}
    \gridline{\fig{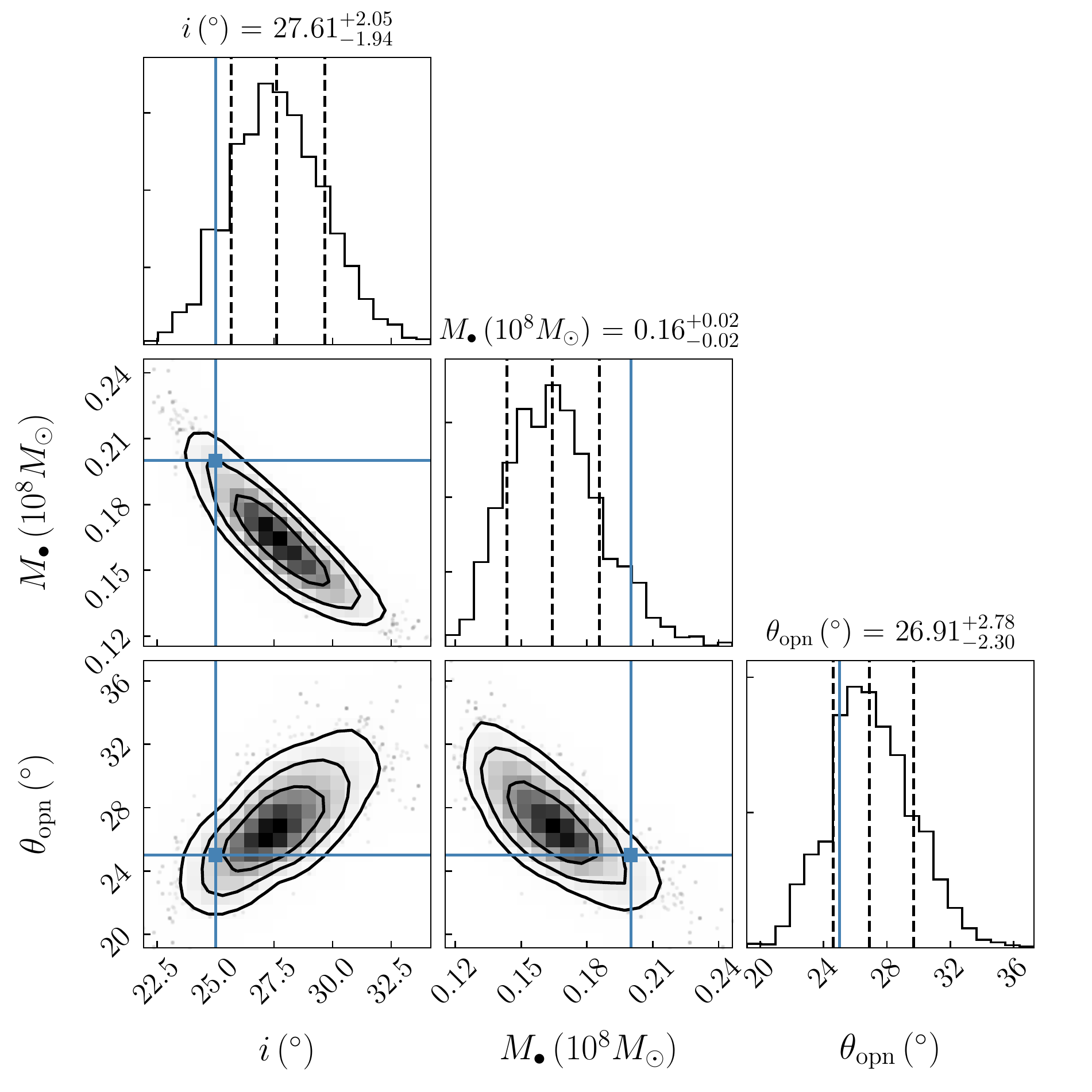}{0.6\textwidth}{}}
    \caption{Probability density distributions of black hole mass $M_{\bullet}
    $, half opening angle $\theta_{\rm opn}$ and inclination angle $i$. The 
    median values of the parameters are given on the tops of panels. Error bars 
    are quoted at the 1$\sigma$ level, which are given by each distribution. 
    The blue lines represent input values. The dashed lines in the 
    one-dimensional distributions are the 16\%, 50\% and 84\% quantiles, and 
    contours are at $1\sigma$, $1.5 \sigma$ and $2\sigma$ respectively. As we 
    can see, strong correlation between $M_{\bullet}$, $i$ and $\theta_{\rm opn}
    $ can lead to large bias when fitting. The figure is plotted using the 
    \texttt{corner} module developed by \cite{corner2016}. \label{fig:contour}}
\end{figure}

As shown in Fig. \ref{fig:sim_error}(b), for $M_{\bullet}$, $i$ and $\theta_
{\rm opn}$, there is no systematic change of uncertainty or bias when error of 
DPC data varies, because these parameters are mainly constrained by profiles of 
emission lines. But we note that there is a dramatic increase of bias for $M_
{\bullet}$ when $\delta_{\rm DPC} = 20\%$. It may be caused by strong 
correlation between $M_{\bullet}$, $i$ and $\theta_{\rm opn}$, as shown in Fig. 
\ref{fig:contour}. The correlation comes from the fact that the FWHM of the 
emission line can remain unchanged if we increase black hole mass and decrease 
opening or inclination angle simultaneously. In the case of strong correlation, 
input values have exceeded $1 \sigma$ range in 1 dimensional distributions, 
though they are still on the boundary of $1 \sigma$ range in 2 dimensional 
distributions. Furthermore, the sampling algorithm we used also performs worse 
when correlations between parameters get stronger. Thus, strong correlation 
between parameters can lead to large bias in fitting.

\subsection{Amplitudes of DPCs}
\begin{figure}
    \plottwo{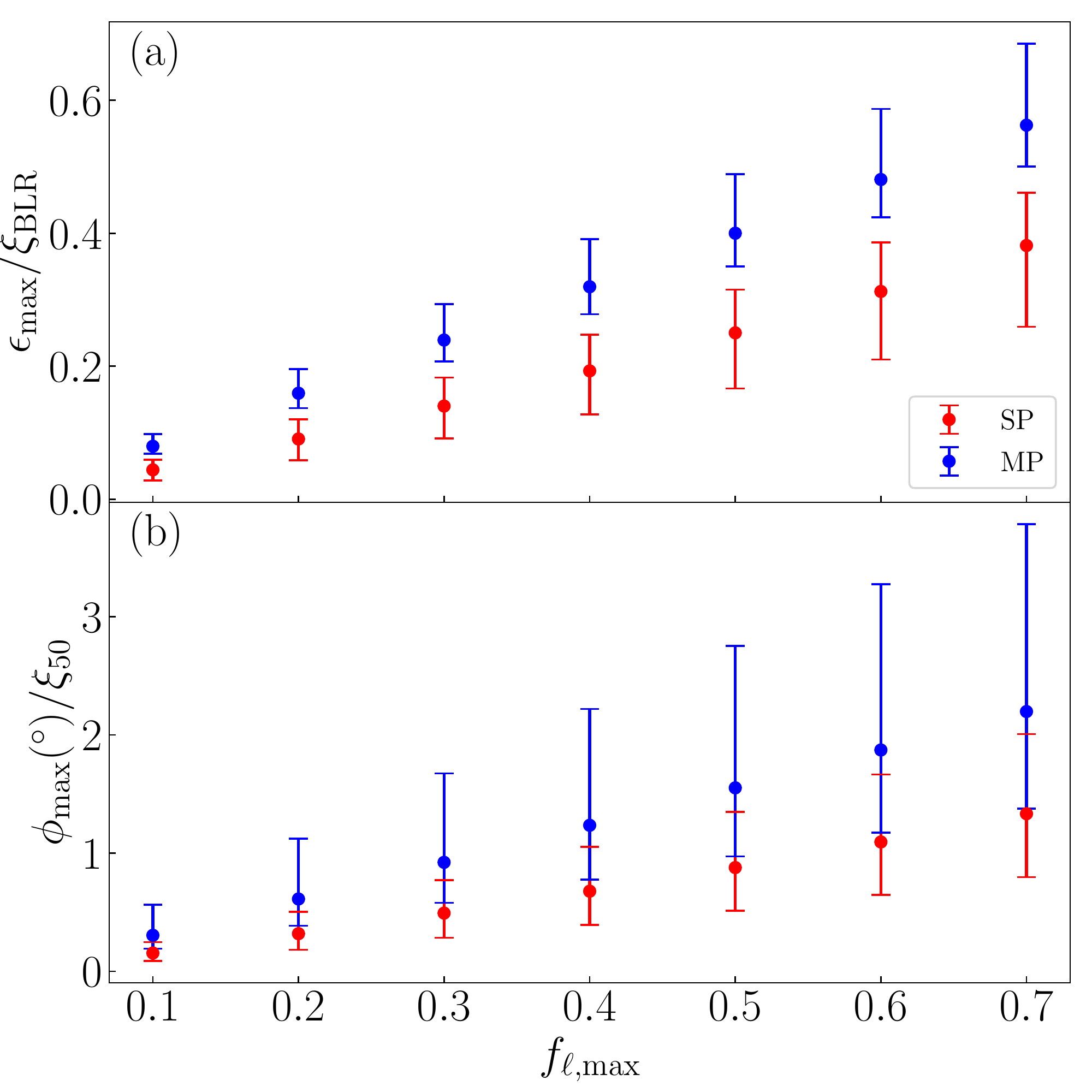}{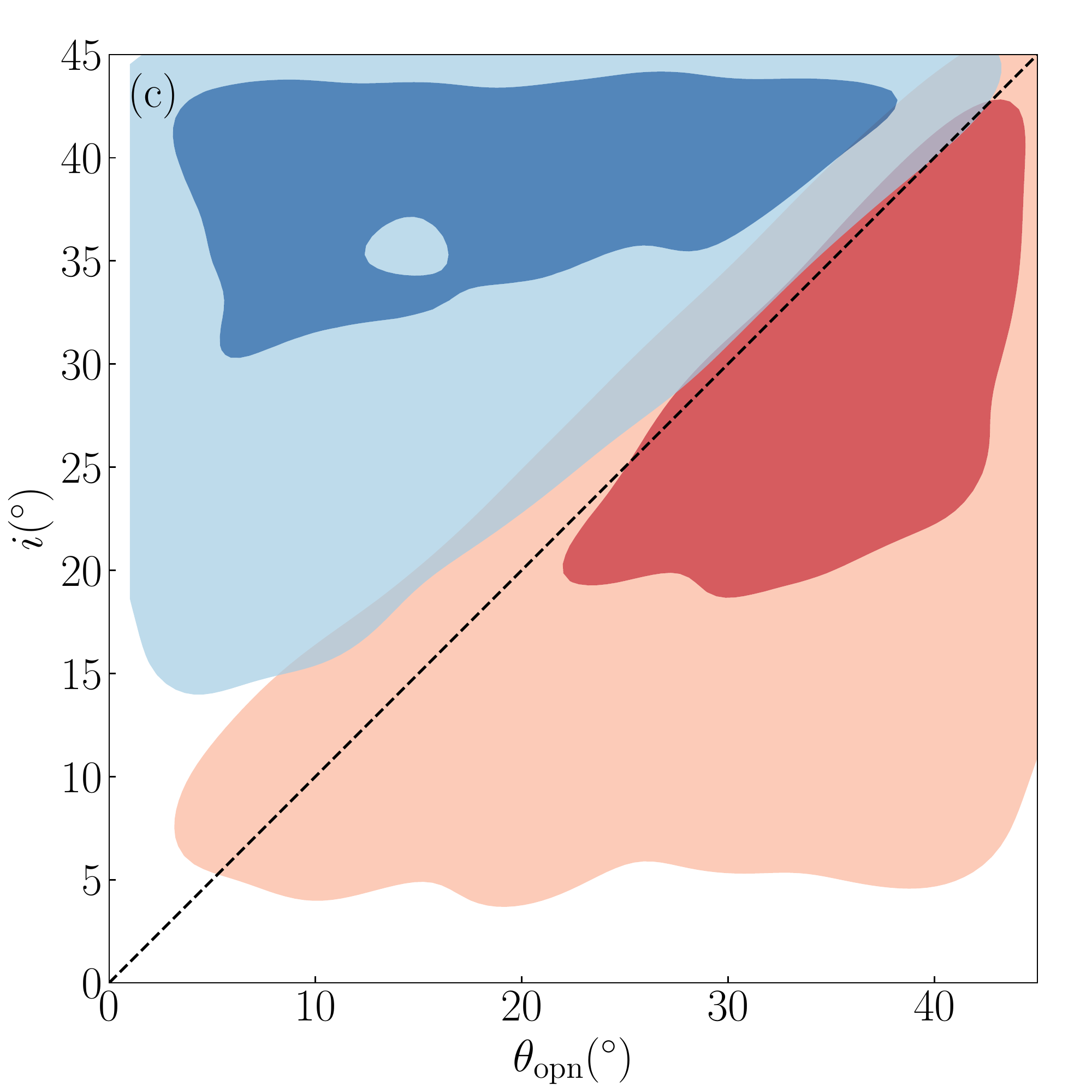}
    \caption{(a)The ratio of photocenter displacement to average angular size 
    of BLR under different emission line strength in our simulated sample. It 
    is evidently that photocenters of targets with multiple peaks in their 
    profiles are lager than those with single peaks. (b) Predicted amplitudes 
    of differential phases observed by GRAVITY. Here, we assume that average 
    angular sizes of BLRs are $50 {\rm\,\mu as}$ ($\xi_{50} = \xi_{\rm BLR} / 50
    {\rm\,\mu as}$) and targets are at the zenith when observed. Ranges of 
    phase amplitudes for targets with multiple peaks in their profiles and 
    those with single peaks partially overlap due to the large uncertainties of 
    position angles. (c) Distribution of inclination and opening angles for 
    objects with single (red region) and multiple (blue region) peaks in 
    profiles respectively. The dark and light color represents $1\sigma$ and 
    $2\sigma$ levels respectively. When $i > \theta_{\rm opn}$, profiles 
    usually possess multiple peaks. \label{fig:signal}}
\end{figure}

In order to estimate the precision of distance measurement for a sample of 
targets, we need to predict amplitudes of DPCs for given magnitudes, redshifts 
and profiles of emission lines. From Eq. \ref{eq:phase}, the differential phase 
depends on the photocenter $\bm{\epsilon}(\lambda)$ of the target as well as 
its projection to the baseline $\bm{B}$ of the interferometer.

Given the redshift of an AGN, its angular and luminosity distance can be 
estimated easily using current values of cosmological parameters. The 
luminosity of the AGN can then be obtained from its magnitude. The widely used 
$R-L$ relation \citep[][]{bentz2013} (but see \cite{du2019} for its revised 
version) now comes in and predicts the average size of the BLR. However, a 
small inclination angle or large opening angle would make the system more 
symmetric, leading to a photocenter displacement much smaller than the average 
angular size of the BLR. Meanwhile, when the emission line is weak compared to 
the continuum, the peak of $\bm{\epsilon}(\lambda)$ will decrease significantly 
from Eq. \ref{eq:fraction}.

In order to quantify all these effects, we choose $i$, $\theta_{\rm opn}$, 
$F$ and $\beta$ randomly from the ranges shown in Table \ref{tab:BLR} to 
generate a large sample of $\bm{\epsilon}_{\ell}(\lambda)$ and corresponding 
line profiles. Dimensional parameters $D_{\rm A}$, $R_{\rm BLR}$ and $M_
{\bullet}$ will be fixed since they can only change the overall amplitude and 
width rather than the shape of the DPC and profile. We divide the sample into 
two categories according to the number of peaks in their line profiles. For 
each category, we calculate the median value and $1 \sigma$ limit of the ratio 
$\epsilon_{\rm max}/\xi_{\rm BLR}$ under different $f_{\rm \ell,max}$, where 
$\xi_{\rm BLR}$ is the average angular size of the BLR, and $\epsilon_{\rm max}
$ and $f_{\ell,max}$ are the maximum value of $|\bm{\epsilon}(\lambda)|$ and $f_
{\ell}(\lambda)$ respectively. The result is shown in Fig. \ref{fig:signal}(a). 
Clearly, the displacement of photocenter is roughly proportional to the ratio 
of line flux to total flux at line center, and photocenters of targets with 
multiple peaks in their profiles are systematically lager than those with 
single peaks.

In order to predict the peak amplitude of actual phase signal, we must know the 
projected base line configuration and position angle of the target. For 
simplicity, we assume targets are at the zenith when observed and the 
configuration of baselines are the same as that of VLTI. Position angles of 
targets are uniformly distributed between $0\degr$ and $360\degr$. The result 
is shown in Fig. \ref{fig:signal}(b). Similarly, phase amplitudes of targets 
with multiple peaks in their profiles are larger than those with a single peak.

\subsection{Impacts of inclination and opening angles}
\begin{figure}
    \gridline{\fig{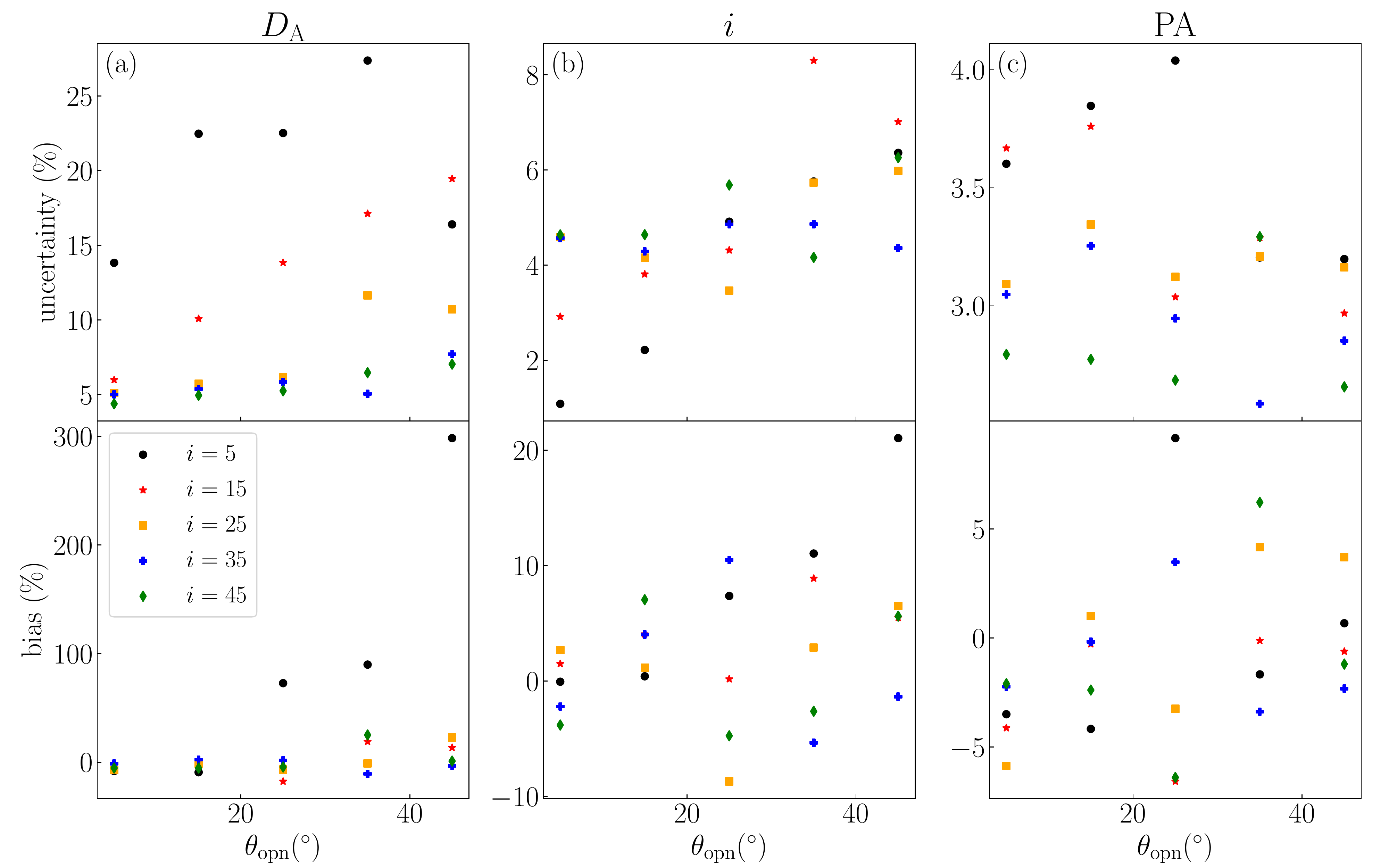}{0.9\textwidth}{}}
    \gridline{\fig{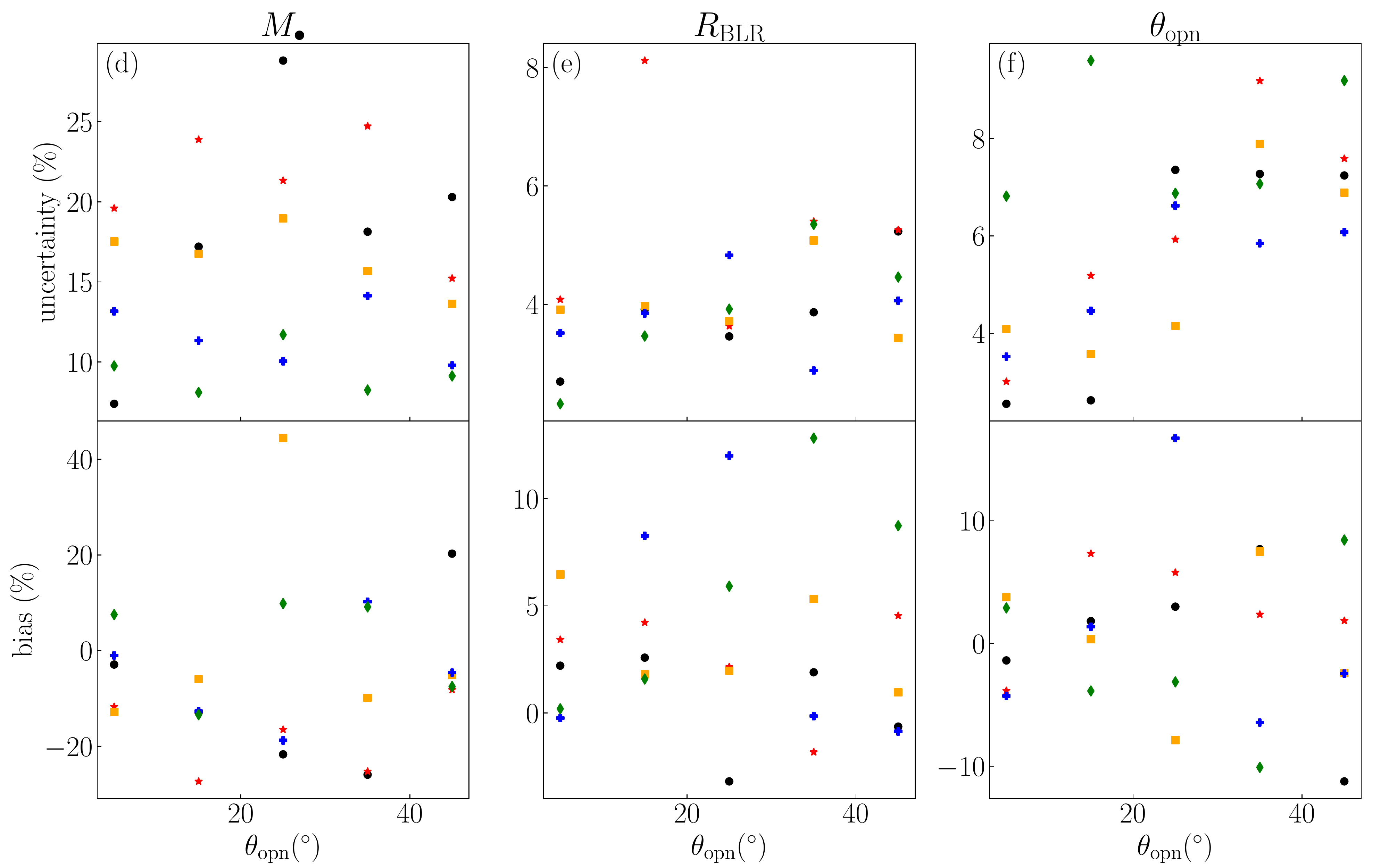}{0.9\textwidth}{}}
    \caption{Relative uncertainties and bias of part of model parameters under 
    different inclinations and opening angles. The mock data of SA and RM are 
    both generated by the BLR model of the same parameter values shown in Table 
    \ref{tab:BLR} except inclinations and opening angles. Relative errors of 
    SARM data are fixed. The configuration of baselines follows the observation 
    of 3C 273. So the target is observed four times for SA data.
    \label{fig:sim_par}}
\end{figure}

Even if the data qualities are the same, uncertainties and accuracies of 
distance measurements can vary with shapes of DPCs and profiles. Dimensional 
parameters $D_{\rm A}$, $R_{\rm BLR}$ and $M_{\bullet}$ can only change the 
overall amplitude and width rather shape of the DPC and profile. Their impacts 
on the distance measurement can be converted to impacts of relative error of 
the data. Among remaining parameters, $i$ and $\theta_{\rm opn}$ can change the 
shape of the DPC and profile drastically \citep{rakshit2015,songsheng2019}, 
thus altering degeneracies among all parameters, consequently further changing 
the uncertainty of parameter inference. In order to study their impacts, we 
vary $i$ and $\theta_{\rm opn}$ from $5\degr$ and $45\degr$ respectively, and 
keep other parameters the same as those in subsection \ref{sec:qualities} and 
relative error of the DPC at $20\%$. For each data set, we perform the Bayesian 
analysis and calculate relative uncertainty and bias for each parameter. The 
results are shown in Fig. \ref{fig:sim_par}.

\begin{figure}
    \gridline{\fig{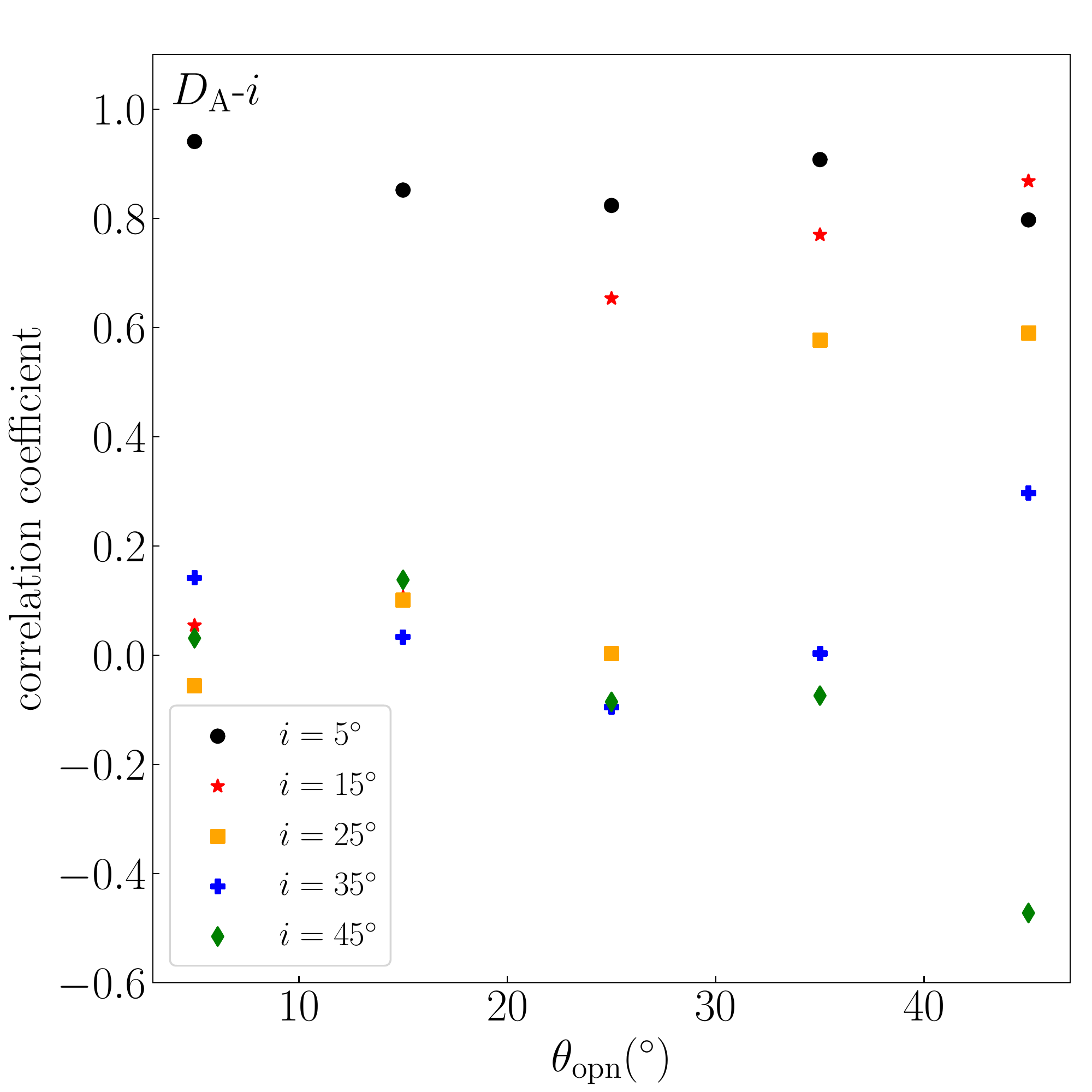}{0.5\textwidth}{}}
    \caption{Pearson product-moment correlation coefficients of angular distances and inclination angles under different inclinations and opening angles. Generally, the smaller the inclination angle, the stronger the correlation between the distance and the inclination angle. For moderate inclination angles ($i \sim 15\degr - 25\degr$), correlations also increase with opening angles of BLR. Correlation coefficients are calculated through \texttt{corrcoef} method in python's \texttt{numpy} module.\label{fig:correlation}}
\end{figure}

Generally, uncertainties of $D_{\rm A}$ becomes large when $i$ decreases, as 
illustrated in Fig. \ref{fig:sim_par}(a). For low inclinations (such as $i 
\lesssim 10\degr$), the profile of broad emission line usually posses a single 
peak (slightly depends on opening angles), as shown in Fig.\ref{fig:signal}(c). 
In this case, the inclination is hard to be determined accurately due to 
degeneracies with other parameters. Furthermore, the LOS approaching the 
symmetry axis will significantly increase the symmetry of the system, thus 
decrease the amplitude of DPC drastically. Thus, the correlation between $i$ 
and $D_{\rm A}$ becomes much stronger, as shown in Fig. \ref{fig:correlation}, 
leading to larger uncertainties of $D_{\rm A}$.

For moderate inclinations ($i \sim 15\degr-25\degr$), uncertainties of $D_{\rm 
A}$ also increase with opening angles of BLRs due to increasing symmetries. 
There are obvious leaps of correlations between $D_{\rm A}$ when $\theta_{\rm 
opn} = 25\degr (i = 15\degr)$ or $\theta_{\rm opn} = 35\degr (i = 25\degr)$, 
since profiles become single-peaked when $\theta_{\rm opn} > i$. If 
inclinations become larger ($i \sim 35\degr-45\degr$), systems will keep 
asymmetry and degeneracies between $D_{\rm A}$ and $i$ are weak, contributing 
to much smaller uncertainties of $D_{\rm A}$.

The variation of the $D_{\rm A}$-bias with inclinations and opening angles 
is roughly similar. We emphasis that the Bayesian inference becomes very 
unstable for a system with low inclinations and large opening angles because 
the system will be highly symmetric in the view of the observer. In such a 
case, extremely large biases $(\ge 100\%)$ appear, even much larger than 
uncertainties, and the inferred value of distance can not be trusted.

In Fig. \ref{fig:sim_par}(b), the evolution of the uncertainty of $i$ 
with opening angles is similar to that of $D_{\rm A}$. There is no systematic 
increase of $\delta_{i}$ as $i$ decreases since our definition of $\delta_{i}$ 
is $\Delta i / {\rm 1\,rad}$ rather than $\Delta i / i$.
In Fig. \ref{fig:sim_par}(c), uncertainties of $\rm PA$ measurements increase 
slightly when inclinations decrease, but they can be determined much more 
accurately compared to other parameters. In Fig. \ref{fig:sim_par}(d), 
uncertainties of measurements of $M_{\bullet}$ also tend to be large for 
face-on targets, since a slight change of $i$ can cause a significant variation 
of width of DPC or profile when inclination is small. In Fig. \ref{fig:sim_par}
(e), inference of $R_{\rm BLR}$ is less affected by the inclination and opening 
angle, as one dimensional RM is insensitive to them. In Fig. \ref{fig:sim_par}
(f), uncertainties of $\theta_{\rm opn}$ vary with inclinations and opening 
angles similarly to those of $i$ due to correlations between $i$ and $\theta_
{\rm opn}$. Generally, values of $\Delta i$ and $\Delta \theta_{\rm opn}$ are 
relatively close.

\section{Discussion}
\subsection{Error budget}
\begin{figure}
    \gridline{\fig{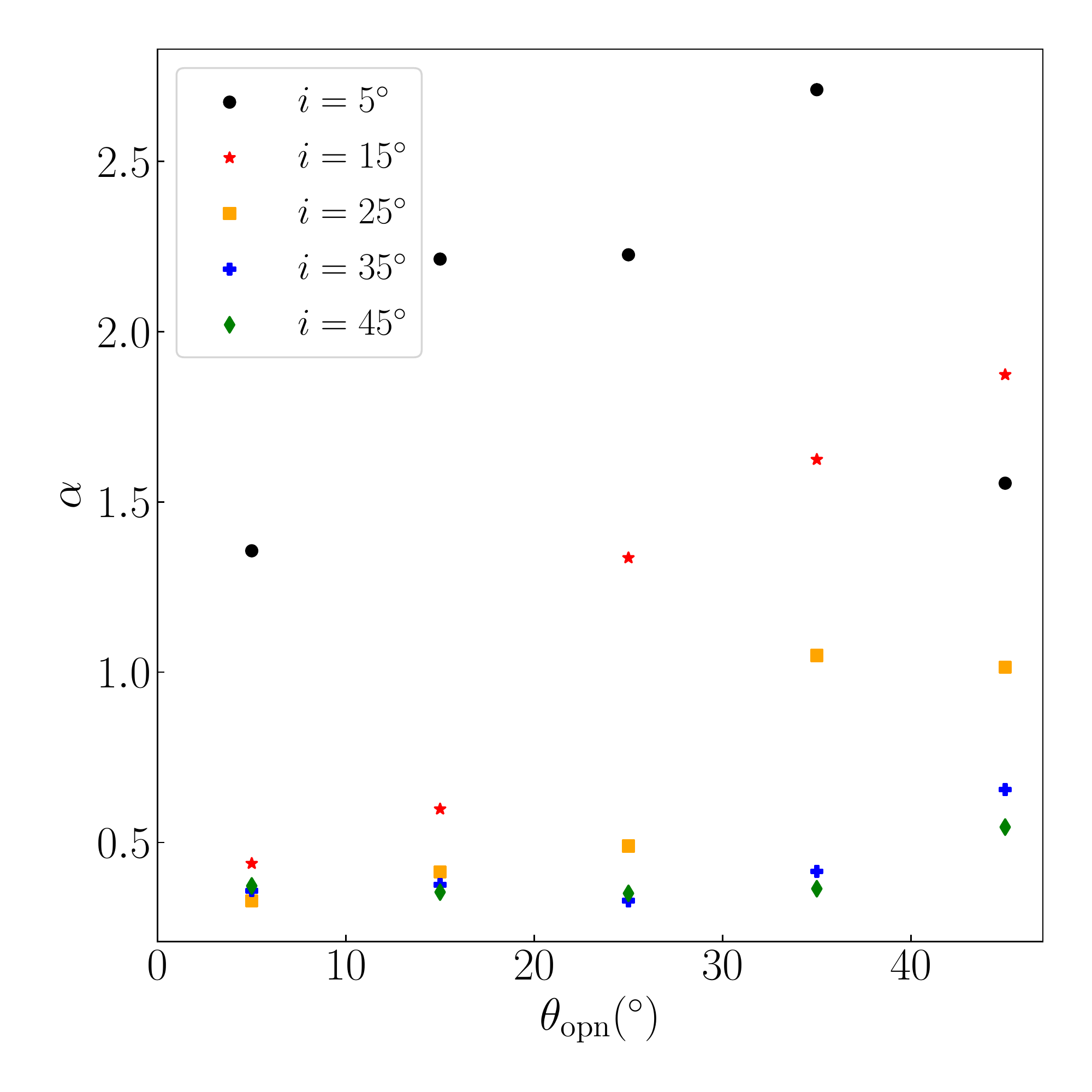}{0.5\textwidth}{}}
    \caption{The ratio between the relative uncertainty of angular size of BLR 
    and that of DPC ($\alpha$) under different inclinations and opening angles.
    For very small inclinations ($i \sim 5\degr$), the relative uncertainty of angular size of BLR is twice that of DPC, while for large inclinations ($i \sim 35\degr-45\degr$), it is less than half of the DPC.
    \label{fig:alpha}}
\end{figure}

The statistical relative uncertainties of distance measurements can be written 
as
\begin{equation}
  \delta_{D_{\rm A}} = \sqrt{\delta_{R_{\rm BLR}}^2 
    + \alpha^2\delta_{\rm DPC}^2},
\end{equation}
where the coefficient $\alpha$ depends on inclination and opening angles of 
BLRs. Multiple observations ($N$) of a target can significantly reduce the 
statistical uncertainty. Considering these influences, we have
\begin{equation}\label{eq:relative_uncertainty}
    \delta_{D_{\rm A}} = \sqrt{\delta_{R_{\rm BLR}}^2 + 
      \alpha^{2}\delta_{\rm DPC}^2/N}.
\end{equation}
From simulations shown in Fig. \ref{fig:sim_par}, we can estimate the value of 
$\alpha$ for different inclination and opening angles \footnote{Note that we 
have $N=4$ in our simulations.}. The result is shown in Fig. \ref
{fig:alpha}. When the inclination is very small ($i \sim 5\degr$), the 
$\alpha$ is around $2$, and the bias of inference is too large to obtain a 
credible measurement. For moderate inclinations, ($i \sim 15\degr-25\degr$), we 
have $\alpha \sim 0.5$ when $i \ga \theta_{\rm opn}$ and $\alpha \sim 1.5$ when 
$i \la \theta_{\rm opn}$. For large inclinations, ($i \sim 35\degr-45\degr$), 
the $\alpha$ is about $0.4$.

Inclinations and opening angles can be in principle inferred through number 
of peaks in line profiles. As shown in Fig \ref{fig:signal}(c), if inclinations 
are large and opening angles are small ($i > \theta_{\rm opn}$), simulated 
profiles usually possess multiple peaks, and vice versa. So selecting objects 
with multiple peaks in their profiles can reduce uncertainties in distance 
measurements notably. In practice, due to the blend of narrow line emission and 
instrument broadening effect, the proportion of objects with multiple peaks in 
profiles in real samples is much smaller than that in our simulated samples 
\citep[e.g., see][Table ~2]{liu2019}. A quick way to estimate inclinations and 
opening angles of a large sample of objects is to compare their profiles with 
template profiles simulated by BLR models. Objects with $i < 10^{\circ}$ or 
$\theta_{\rm opn} > 40^{\circ}$ can be excluded at first step. Before making 
observation plans, we can fit profiles of specific objects to the BLR model to 
obtain a reliable estimation for inclination and opening angles \citep[e.g.,][]
{raimundo2019,raimundo2020}. Then we can predict the expected DPC for each 
object and choose appropriate ones for SARM campaign. Other ways to determine 
inclinations of AGNs includes narrow line region kinematics \citep{fischer2013} 
and pc-scale radio jet \citep[e.g.][]{kun2015}.

Systematic errors of distance measurement through SARM have been discussed 
thoroughly in \cite{wang2020}. Firstly, RM campaign measures the region of 
optical emission line (usually $\rm H\beta$ for objects with low redshifts) 
while SA conducted by GRAVITY can only measure the region of NIR emission line 
(usually $\rm Br \gamma$ or $\rm Pa \alpha$). Their size may be sightly 
different due to the different optical depths. But this can be reduced by 
comparing widths of different emission lines in joint analysis of 
velocity-resolved RM and SA observations. For 3C 273, the difference between 
the size of $\rm H\beta$ and $\rm Pa \alpha$ emitting region is estimated to be 
$13\%$, which is slightly smaller than the statistical error.  A RM campaign 
using the same emission line as that in SA observation are highly needed to 
quantify this effect. 
%
Secondly, RM campaign measures the variable part of a BLR 
while SA observes the entire region, resulting in systematic errors in the SARM 
analysis. By comparing the shape and width of mean with root-mean-square (RMS) 
spectra across the whole RM campaign, the error can also been assessed. For 3C 
273, the variable part of its BLR show little difference with the entire BLR. 
{\cblue Finally, signatures of inflow or outflow has been found recently by 
analysing high-quality RM data \citep{pancoast2014b,grier2013,xiao2018}.
The general shape of differential phase curve for an inflow BLR is similar to 
that of a Keplerian BLR, except that the displacement of photocenter would be 
parallel to the axis rather than perpendicular to it \citep{rakshit2015}.
The angular size of BLR can still be well constrained by SA observation, and so 
the uncertainty of distance through SARM analysis will not change a lot. 
However, the relation between clouds' velocities and black hole mass will be 
much more ambiguous, increasing the uncertainty of black hole mass measurement.}
Based on these limited information and the fact that angular distance of 3C 273 
measured by SARM is consistent with that from the current cosmological model, 
we draw a conclusion that systematic errors are at most comparable to 
statistical errors.

\subsection{Model tests}
In our mock data analysis, the correctness of BLR model is always guaranteed 
and the all fittings can reach $\chi^2 \approx 1$ level. However, BLRs are 
diverse individually. Our simplest model works quite well for 3C 273 since its 
profile of emission line are symmetric. When selecting targets, objects with 
simple and regular line profiles should be prioritized. The response of 
emission line to continuum also needs to be clear to avoid complications by 
long term trending, BLR "holiday", multiple lags or fast breathing. Before 
Bayesian analysis, model-independent methods, such as reconstruction of 
velocity-delay map through maximum entropy method or calculation of centroid 
position of the photocenter in several wavelength channels, should be applied 
to obtain the basic geometry and dynamics of the BLR. The appropriate model 
with minimum number of parameters will be tested first. Subsequently, more can 
be added until a good fitting reaches.

Actually, radial distribution of BLR clouds described by Eq. (\ref{eq:gamma1}, 
\ref{eq:gamma2}) can be alternatively assumed by other forms, such as Gaussian 
distributions \citep{pancoast2011}. This leads to different results of 
distances from the fittings, which could be regarded as one of systematical 
errors. This can be addressed by evaluating the Bayesian evidence of the model, 
which is a prime result of nested sampling method \citep{john2006,shaw2007}. 
Moreover, SMBH mass can be simultaneously obtained by the SARM analysis. We 
will discuss these issues in an separate paper.

\subsection{$H_{0}$-tension}
\begin{figure}
    \plotone{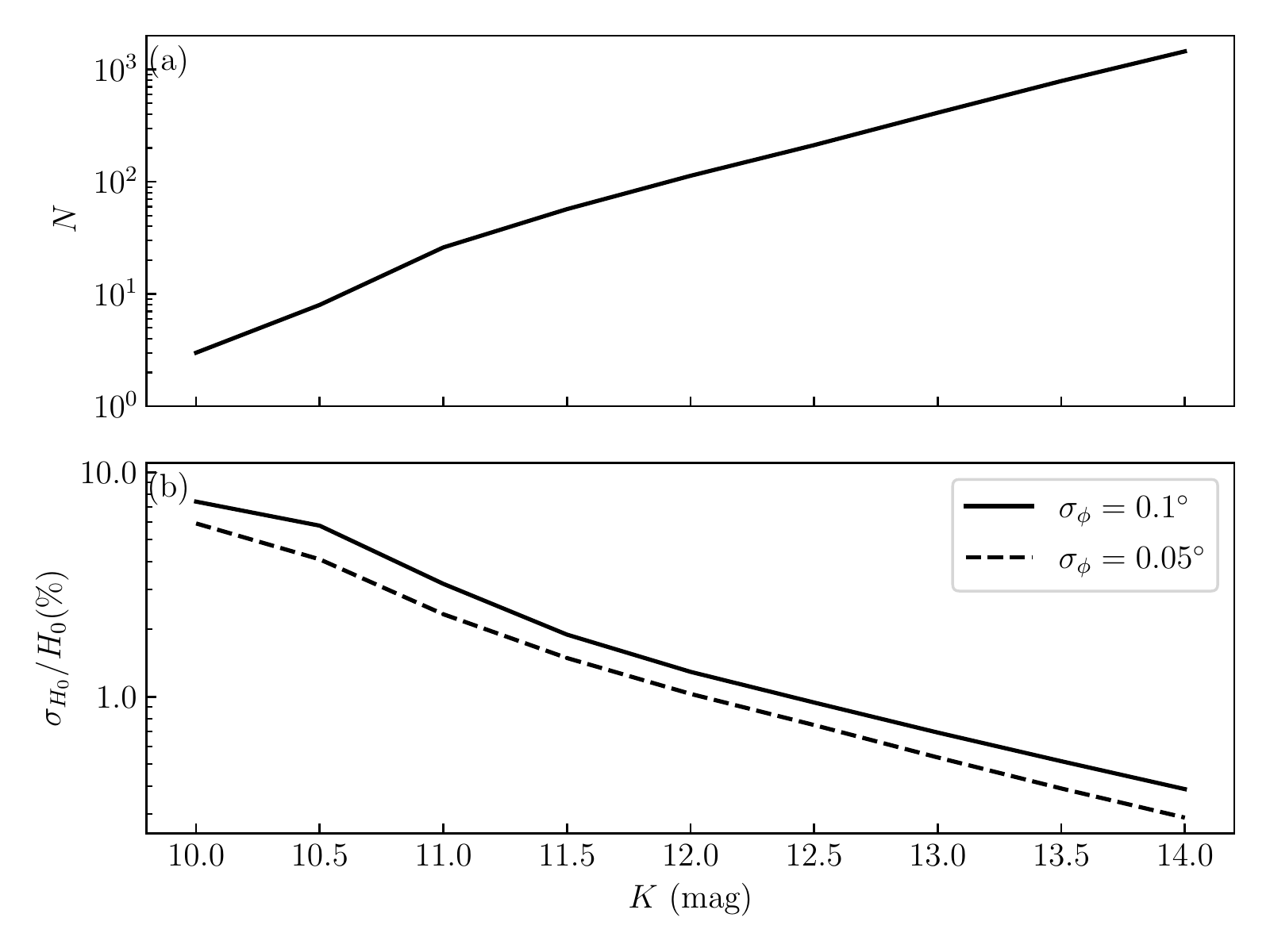}
    \caption{(a) The expected number of Type I AGNs which can be observed using 
    GRAVITY as the limiting $K$ band magnitude increases. 
    (b) The corresponding $H_0$ measurement error through SARM, assuming each 
    object are observed four times. $\sigma_{\phi}$ is the uncertainties of 
    phase in each observation for each base line. The uncertainties of linear 
    size of BLR through RM are kept at $10\%$. Systematical error bars risen 
    from the different lines between RM and SA observations can be inprinciple
    alleviated by either mapping the same line with SA or through kinematic
    relation of the SARM lines. 
    \label{fig:cosmo}}
\end{figure}

Since the beginning of the last century, advances in the distance measurement 
of extragalactic objects have been driving the development of cosmology. In 
particular, the cosmic distance ladder using Cepheid and Type Ia supernovae as 
standard candles has made great achievements, including the discovery of the 
expansion of the universe \citep{hubble1929} as well as its acceleration \citep
{riess1998,perlmutter1999}. Currently, the Equation of State of Dark energy 
(SH0ES) program achieves a measurement of the Hubble constant ($H_0$) with $1.
91\%$ precision based on the distance ladder, providing a value of $H_0 = 74.03 
\pm 1.42 {\rm\,km\,s^{-1}\, Mpc^{-1}}$ \citep{riess2019}. However, observations 
of the cosmic microwave background (CMB) radiation from the early universe 
enables an indirect inference of $H_0$ under the assumption of the flat 
$\Lambda$CDM cosmological model, giving $H_0 = 67.4 \pm 0.5 {\rm\,km\,s^{-1} \,
Mpc^{-1}}$ \citep{planck2018}. Tensions between the early and late Universe 
measurements are at the level of $\sim 4.4 \sigma$ \citep{verde2019}, 
indicating that either unaccounted systematic errors might bias at least one of 
these measurements \citep[see discussion of its possibility in][]{davis2019,
rameez2019,rose2020}, or the flat $\Lambda$CDM cosmological model needs to be 
extended to include new physics, such as dynamical/interacting dark energy 
\citep[e.g.,][]{divalentino2016, divalentino2017, huang2016,zhao2017}, dark 
radiation \citep[e.g.,][] {buenabad2015, ko2016}, neutrino self-interaction 
\citep[e.g.,][]{blinov2019, kreisch2020}, and so on.

At such a cross roads, new approaches to determining Hubble constant without 
calibration of the local distance ladders or assumption of the flat 
$\Lambda$CDM cosmological model are particularly important as arbitrators. 
Gravitational wave emitted by binary neutron star merger can be used as 
``standard siren'' to determine the luminosity distance of the binary \cite 
{schutz1986}. The recent observation of the merger signal GW180817 \citep
{abbott2017} along with its optical counterpart \citep{abbott2017b,
soaressantos2017} provided the first standard siren measurement of $H_0$ with 
precision $14\%$ \citep {abbott2017c}. Its constraint ability of cosmological 
parameters, including Hubble constant, has been extensively studied \citep
{cai2017,chen2018}. Another promising approach relies on strong gravitational 
lenses with measured time delays between the multiple images \citep{refsdal1964}. 
The program $H_0$ Lenses in COSMOGRAIL's Wellspring (H0LiCOW) \citep
{suyu2017} obtained a $2.4\%$ measurement of $H_0$ from a joint analysis of six 
gravitationally lensed quasars, in agreement with measurements of $H_0$ by the 
local distance ladder \citep{wong2020}.

Systematic errors caused by BLR size discrepancy of different emission lines
are not included in Fig \ref{fig:cosmo}, making it an optimistic estimation of 
$H_0$ constraints by SARM project. However, if we use the velocity resolved RM 
data by including profiles of optical emission lines, BLR sizes of optical and 
NIR lines can be different in the model. The systematic errors can be
alleviated.  

In order to study the constraint ability on $H_0$ by joint analysis of SARM 
data, we generate a mock sample of type I AGNs for GRAVITY/GRAVITY+. Details on 
the generation and properties of the sample are presented in Appendix \ref
{sec:mock sample}. For each object in the sample, the profile and DPC are 
simulated using BLR models. Once we obtain the maximum amplitudes of phase 
curves $\phi_{\rm max}$, uncertainties of distance measurements for each object 
can be estimated by Eq. \ref{eq:relative_uncertainty}. If other cosmological  
parameters are fixed except $H_0$, the uncertainty of $H_0$ measurement using 
the whole sample can be obtained. 

{\cblue 
Since 200 day light curves with 1 day cadence and no gaps can hardly be 
achieved in typical RM campaigns, we would assume the relative uncertainty of 
BLR size measurement through RM campaign is $10\%$ rather than $2\%$ shown in 
Fig. \ref{fig:sim_error}(a). (e.g., a recent RM campaign conducted by \cite
{bentz2021} for NGC 3783, which is also a suitable target for GRAVITY, obtained 
a time lag measurement with $10\%$ uncertainty.)}
We also assume the uncertainty of phase measurement for each baseline is $0.
1\degr$. Note that highly face-on objects ($i \lesssim 10\degr$) are discarded 
due to their large biases in distance measurements. 
\footnote{
    \cblue If orientations of BLRs are randomly distributed and the largest 
    inclination of Type I AGN is $45\degr$, fraction of objects with $i\lesssim 
    10\degr$ is less than $5\%$.}
Objects with maximum phases less than $0.1 \degr$ are also excluded since they 
help little to constrain the Hubble constant.The result is shown in Fig \ref
{fig:cosmo}. As we can see, by observing $60$ targets with $K$ band magnitudes 
better than $11.5$ and maximum phase larger than $0.1\degr$ through GRAVITY, 
the uncertainty of $H_0$ will be less than $2\%$. Currently, GRAVITY can 
achieve a phase error of $0.1 \degr$ per baseline in observations of bright ($K 
\sim 10-11$) AGNs \citep{gravity2018,gravity2020,dexter2020}. But the 
integration time for each object may reach 10 hours due to the limit of fringe 
tracking. Fortunately, GRAVITY+\footnote{see its detail information from https:/
/www.mpe.mpg.de/ir/gravityplus}, the upgraded version of GRAVITY, aims to 
achieve on-axis fringe tracking for AGNs as faint as $K \sim 14-15$ in the near 
future \citep{dexter2020}. If we assume that GRAVITY+ requires $1-2$ hour of 
integration time to reach a phase error of $0.1 \degr$ per baseline for these 
bright targets, $60-120$ hours of observation time can measure the Hubble 
constant with better than $2\%$ accuracy.

Finally, we would like to point out that a sample composed of 53 AGNs for 
future SARM campaigns has been selected by \cite{wang2020} from the current
catalogs (mainly from Veron-Cetty and Verson's catalog, and 2dF and 6dF ect). 
AGNs surveyed by 4MOST\footnote{https://www.4most.eu/cms/} in future will 
be greatly increased in the southern hemisphere, moreover powerful 
GRAVITY+ onboard VLTI will conveniently observe fainter AGNs and quasars at 
cosmic noon between $z=2-3$, offering opportunities of measuring cosmic 
expansion history from local Univerise to the deeper. 

\section{Conclusion}
In this paper, we conduct a mock data analysis for cosmology through SARM  
campaign. We have tested how the relative uncertainties of distance 
measurements depend on errors of DPCs as well as the roles of inclinations and 
opening angles of broad-line region (BLR). 
For BLRs with inclinations $\gtrsim 10\degr$ and opening angles $\lesssim 
40\degr$, analyses of SARM data can generate reliable quasar distances even for
relatively poor SA measurements with relative error of $40\%$ for GRAVITY-like 
facilities. If the limiting magnitude of the GRAVITY reaches $11.5$ in $K$ band 
and errors of phase measurements are as low as $0.1\degr$, the SARM campaign 
can constrain $H_0$ to an uncertainty of $2\%$ by observing $60$ targets.

\acknowledgments
We thank Jinyi Shangguan for useful comments on the mansucript. We are also grateful to the members of the IHEP AGN group for enlightening dicsussions.
JMW thanks the support by National Key R\&D Program of China through grant -2016YFA0400701, 
by NSFC through grants NSFC-11991050, -11991054, -11833008, -11690024, and by grant 
No. QYZDJ-SSW-SLH007 and No.XDB23010400.

\newpage
\newcommand{\noop}[1]{}

\begin{thebibliography}{}
    \expandafter\ifx\csname natexlab\endcsname\relax\def\natexlab#1{#1}\fi
    \providecommand{\url}[1]{\href{#1}{#1}}
    \providecommand{\dodoi}[1]{doi:~\href{http://doi.org/#1}{\nolinkurl{#1}}}
    \providecommand{\doeprint}[1]{\href{http://ascl.net/#1}{\nolinkurl{http://ascl.net/#1}}}
    \providecommand{\doarXiv}[1]{\href{https://arxiv.org/abs/#1}{\nolinkurl{https://arxiv.org/abs/#1}}}
    
    \bibitem[{{Abbott} {et~al.}(2017{\natexlab{a}}){Abbott}, {Abbott}, {Abbott},
      {Acernese}, {Ackley}, {Adams}, {Adams}, {Addesso}, {Adhikari}, {Adya},
      {Affeldt}, {Afrough}, {Agarwal}, {Agathos}, {Agatsuma}, {Aggarwal}, {Aguiar},
      {Aiello}, {Ain}, {Ajith}, {Allen}, {Allen}, \& {Allocca}}]{abbott2017}
    {Abbott}, B.~P., {Abbott}, R., {Abbott}, T.~D., {et~al.} 2017{\natexlab{a}},
      \prl, 119, 161101, \dodoi{10.1103/PhysRevLett.119.161101}
    
    \bibitem[{{Abbott} {et~al.}(2017{\natexlab{b}}){Abbott}, {Abbott}, {Abbott},
      {Acernese}, {Ackley}, {Adams}, {Adams}, {Addesso}, {Adhikari}, {Adya},
      {Affeldt}, {Afrough}, {Agarwal}, {Agathos}, {Agatsuma}, {Aggarwal}, {Aguiar},
      {Aiello}, {Ain}, {Ajith}, {Allen}, {Allen}, \& {Allocca}}]{abbott2017b}
    ---. 2017{\natexlab{b}}, \apjl, 848, L12, \dodoi{10.3847/2041-8213/aa91c9}
    
    \bibitem[{{Abbott} {et~al.}(2017{\natexlab{c}}){Abbott}, {Abbott}, {Abbott},
      {Acernese}, {Ackley}, {Adams}, {Adams}, {Addesso}, {Adhikari}, {Adya},
      {Affeldt}, {Afrough}, {Agarwal}, {Agathos}, {Agatsuma}, {Aggarwal}, {Aguiar},
      {Aiello}, {Ain}, {Ajith}, {Allen}, {Allen}, \& {Allocca}}]{abbott2017c}
    ---. 2017{\natexlab{c}}, \nat, 551, 85, \dodoi{10.1038/nature24471}
    
    \bibitem[{{Bentz} {et~al.}(2021){Bentz}, {Street}, {Onken}, \&
      {Valluri}}]{bentz2021}
    {Bentz}, M.~C., {Street}, R., {Onken}, C.~A., \& {Valluri}, M. 2021, \apj, 906,
      50, \dodoi{10.3847/1538-4357/abccd4}
    
    \bibitem[{{Bentz} {et~al.}(2010){Bentz}, {Walsh}, {Barth}, {Yoshii}, {Woo},
      {Wang}, {Treu}, {Thornton}, {Street}, {Steele}, {Silverman}, {Serduke},
      {Sakata}, {Minezaki}, {Malkan}, {Li}, {Lee}, {Hiner}, {Hidas}, {Greene},
      {Gates}, {Ganeshalingam}, {Filippenko}, {Canalizo}, {Bennert}, \&
      {Baliber}}]{bentz2010}
    {Bentz}, M.~C., {Walsh}, J.~L., {Barth}, A.~J., {et~al.} 2010, \apj, 716, 993,
      \dodoi{10.1088/0004-637X/716/2/993}
    
    \bibitem[{{Bentz} {et~al.}(2013){Bentz}, {Denney}, {Grier}, {Barth},
      {Peterson}, {Vestergaard}, {Bennert}, {Canalizo}, {De Rosa}, {Filippenko},
      {Gates}, {Greene}, {Li}, {Malkan}, {Pogge}, {Stern}, {Treu}, \&
      {Woo}}]{bentz2013}
    {Bentz}, M.~C., {Denney}, K.~D., {Grier}, C.~J., {et~al.} 2013, \apj, 767, 149,
      \dodoi{10.1088/0004-637X/767/2/149}
    
    \bibitem[{{Blandford} \& {McKee}(1982)}]{blandford1982}
    {Blandford}, R.~D., \& {McKee}, C.~F. 1982, \apj, 255, 419,
      \dodoi{10.1086/159843}
    
    \bibitem[{{Blinov} {et~al.}(2019){Blinov}, {Kelly}, {Krnjaic}, \&
      {McDermott}}]{blinov2019}
    {Blinov}, N., {Kelly}, K.~J., {Krnjaic}, G., \& {McDermott}, S.~D. 2019, \prl,
      123, 191102, \dodoi{10.1103/PhysRevLett.123.191102}
    
    \bibitem[{Brewer \& Foreman-Mackey(2018)}]{brewer2018}
    Brewer, B.~J., \& Foreman-Mackey, D. 2018, JOURNAL OF STATISTICAL SOFTWARE, 86,
      1, \dodoi{10.18637/jss.v086.i07}
    
    \bibitem[{{Brewer} {et~al.}(2011){Brewer}, {P\'{a}rtay}, \&
      {Cs\'{a}nyi}}]{brewer2011}
    {Brewer}, B.~J., {P\'{a}rtay}, L.~B., \& {Cs\'{a}nyi}, G. 2011, Statistics and
      Computing, 21, 649, \dodoi{10.1007/s11222-010-9198-8}
    
    \bibitem[{{Brotherton} {et~al.}(2020){Brotherton}, {Du}, {Xiao}, {Bao}, {Zhao},
      {McLane}, {Olson}, {Wang}, {Huang}, {Hu}, {Kasper}, {Chick}, {Nguyen},
      {Maithil}, {Hand}, {Li}, {Ho}, {Bai}, {Bian}, \& {Wang}}]{brotherton2020}
    {Brotherton}, M.~S., {Du}, P., {Xiao}, M., {et~al.} 2020, arXiv e-prints,
      arXiv:2011.05902.
    \newblock \doarXiv{2011.05902}
    
    \bibitem[{{Buen-Abad} {et~al.}(2015){Buen-Abad}, {Marques-Tavares}, \&
      {Schmaltz}}]{buenabad2015}
    {Buen-Abad}, M.~A., {Marques-Tavares}, G., \& {Schmaltz}, M. 2015, \prd, 92,
      023531, \dodoi{10.1103/PhysRevD.92.023531}
    
    \bibitem[{{Cai} \& {Yang}(2017)}]{cai2017}
    {Cai}, R.-G., \& {Yang}, T. 2017, \prd, 95, 044024,
      \dodoi{10.1103/PhysRevD.95.044024}
    
    \bibitem[{{Chen} {et~al.}(2018){Chen}, {Fishbach}, \& {Holz}}]{chen2018}
    {Chen}, H.-Y., {Fishbach}, M., \& {Holz}, D.~E. 2018, \nat, 562, 545,
      \dodoi{10.1038/s41586-018-0606-0}
    
    \bibitem[{{Davis} {et~al.}(2019){Davis}, {Hinton}, {Howlett}, \&
      {Calcino}}]{davis2019}
    {Davis}, T.~M., {Hinton}, S.~R., {Howlett}, C., \& {Calcino}, J. 2019, \mnras,
      490, 2948, \dodoi{10.1093/mnras/stz2652}
    
    \bibitem[{{Dexter} {et~al.}(2020){Dexter}, {Lutz}, {Shimizu}, {Shangguan},
      {Davies}, {de Zeeuw}, {Sturm}, {Eisenhauer}, {F{\"o}rster-Schreiber}, {Gao},
      {Genzel}, {Gillessen}, {Pfuhl}, {Tacconi}, \& {Widmann}}]{dexter2020}
    {Dexter}, J., {Lutz}, D., {Shimizu}, T.~T., {et~al.} 2020, arXiv e-prints,
      arXiv:2010.09735.
    \newblock \doarXiv{2010.09735}
    
    \bibitem[{{Di Valentino} {et~al.}(2017){Di Valentino}, {Melchiorri}, \&
      {Mena}}]{divalentino2017}
    {Di Valentino}, E., {Melchiorri}, A., \& {Mena}, O. 2017, \prd, 96, 043503,
      \dodoi{10.1103/PhysRevD.96.043503}
    
    \bibitem[{{Di Valentino} {et~al.}(2016){Di Valentino}, {Melchiorri}, \&
      {Silk}}]{divalentino2016}
    {Di Valentino}, E., {Melchiorri}, A., \& {Silk}, J. 2016, Physics Letters B,
      761, 242, \dodoi{10.1016/j.physletb.2016.08.043}
    
    \bibitem[{{Du} \& {Wang}(2019)}]{du2019}
    {Du}, P., \& {Wang}, J.-M. 2019, \apj, 886, 42,
      \dodoi{10.3847/1538-4357/ab4908}
    
    \bibitem[{{Du} {et~al.}(2016){Du}, {Lu}, {Hu}, {Qiu}, {Li}, {Huang}, {Wang},
      {Bai}, {Bian}, {Yuan}, {Ho}, {Wang}, \& {SEAMBH Collaboration}}]{du2016}
    {Du}, P., {Lu}, K.-X., {Hu}, C., {et~al.} 2016, \apj, 820, 27,
      \dodoi{10.3847/0004-637X/820/1/27}
    
    \bibitem[{{Du} {et~al.}(2018){Du}, {Zhang}, {Wang}, {Huang}, {Zhang}, {Lu},
      {Hu}, {Li}, {Bai}, {Bian}, {Yuan}, {Ho}, {Wang}, \& {SEAMBH
      Collaboration}}]{du2018}
    {Du}, P., {Zhang}, Z.-X., {Wang}, K., {et~al.} 2018, \apj, 856, 6,
      \dodoi{10.3847/1538-4357/aaae6b}
    
    \bibitem[{{Eisenhauer} {et~al.}(2008){Eisenhauer}, {Perrin}, {Brandner},
      {Straubmeier}, {Richichi}, {Gillessen}, {Berger}, {Hippler}, {Eckart},
      {Sch{\"o}ller}, {Rabien}, {Cassaing}, {Lenzen}, {Thiel}, {Cl{\'e}net},
      {Ramos}, {Kellner}, {F{\'e}dou}, {Baumeister}, {Hofmann}, {Gendron}, {Boehm},
      {Bartko}, {Haubois}, {Klein}, {Dodds-Eden}, {Houairi}, {Hormuth},
      {Gr{\"a}ter}, {Jocou}, {Naranjo}, {Genzel}, {Kervella}, {Henning}, {Hamaus},
      {Lacour}, {Neumann}, {Haug}, {Malbet}, {Laun}, {Kolmeder}, {Paumard},
      {Rohloff}, {Pfuhl}, {Perraut}, {Ziegleder}, {Rouan}, \&
      {Rousset}}]{eisenhauer2008}
    {Eisenhauer}, F., {Perrin}, G., {Brandner}, W., {et~al.} 2008, in Society of
      Photo-Optical Instrumentation Engineers (SPIE) Conference Series, Vol. 7013,
      \procspie, 70132A
    
    \bibitem[{{Feroz} {et~al.}(2009){Feroz}, {Hobson}, \& {Bridges}}]{feroz2009}
    {Feroz}, F., {Hobson}, M.~P., \& {Bridges}, M. 2009, \mnras, 398, 1601,
      \dodoi{10.1111/j.1365-2966.2009.14548.x}
    
    \bibitem[{{Fischer} {et~al.}(2013){Fischer}, {Crenshaw}, {Kraemer}, \&
      {Schmitt}}]{fischer2013}
    {Fischer}, T.~C., {Crenshaw}, D.~M., {Kraemer}, S.~B., \& {Schmitt}, H.~R.
      2013, \apjs, 209, 1, \dodoi{10.1088/0067-0049/209/1/1}
    
    \bibitem[{{Fonseca Alvarez} {et~al.}(2020){Fonseca Alvarez}, {Trump},
      {Homayouni}, {Grier}, {Shen}, {Horne}, {Li}, {Brandt}, {Ho}, {Peterson}, \&
      {Schneider}}]{fonseca2020}
    {Fonseca Alvarez}, G., {Trump}, J.~R., {Homayouni}, Y., {et~al.} 2020, \apj,
      899, 73, \dodoi{10.3847/1538-4357/aba001}
    
    \bibitem[{Foreman-Mackey(2016)}]{corner2016}
    Foreman-Mackey, D. 2016, The Journal of Open Source Software, 1, 24,
      \dodoi{10.21105/joss.00024}
    
    \bibitem[{{Gravity Collaboration} {et~al.}(2017){Gravity Collaboration},
      {Abuter}, {Accardo}, {Amorim}, {Anugu}, {{\'A}vila}, {Azouaoui}, {Benisty},
      {Berger}, {Blind}, {Bonnet}, {Bourget}, {Brandner}, {Brast}, {Buron},
      {Burtscher}, {Cassaing}, {Chapron}, {Choquet}, {Cl{\'e}net}, {Collin},
      {Coud{\'e} Du Foresto}, {de Wit}, {de Zeeuw}, {Deen},
      {Delplancke-Str{\"o}bele}, {Dembet}, {Derie}, {Dexter}, {Duvert}, {Ebert},
      {Eckart}, {Eisenhauer}, {Esselborn}, {F{\'e}dou}, {Finger}, {Garcia}, {Garcia
      Dabo}, {Garcia Lopez}, {Gendron}, {Genzel}, {Gillessen}, {Gonte}, {Gordo},
      {Grould}, {Gr{\"o}zinger}, {Guieu}, {Haguenauer}, {Hans}, {Haubois}, {Haug},
      {Haussmann}, {Henning}, {Hippler}, {Horrobin}, {Huber}, {Hubert}, {Hubin},
      {Hummel}, {Jakob}, {Janssen}, {Jochum}, {Jocou}, {Kaufer}, {Kellner},
      {Kendrew}, {Kern}, {Kervella}, {Kiekebusch}, {Klein}, {Kok}, {Kolb}, {Kulas},
      {Lacour}, {Lapeyr{\`e}re}, {Lazareff}, {Le Bouquin}, {L{\`e}na}, {Lenzen},
      {L{\'e}v{\^e}que}, {Lippa}, {Magnard}, {Mehrgan}, {Mellein}, {M{\'e}rand},
      {Moreno-Ventas}, {Moulin}, {M{\"u}ller}, {M{\"u}ller}, {Neumann}, {Oberti},
      {Ott}, {Pallanca}, {Panduro}, {Pasquini}, {Paumard}, {Percheron}, {Perraut},
      {Perrin}, {Pfl{\"u}ger}, {Pfuhl}, {Phan Duc}, {Plewa}, {Popovic}, {Rabien},
      {Ram{\'\i}rez}, {Ramos}, {Rau}, {Riquelme}, {Rohloff}, {Rousset},
      {Sanchez-Bermudez}, {Scheithauer}, {Sch{\"o}ller}, {Schuhler}, {Spyromilio},
      {Straubmeier}, {Sturm}, {Suarez}, {Tristram}, {Ventura}, {Vincent},
      {Waisberg}, {Wank}, {Weber}, {Wieprecht}, {Wiest}, {Wiezorrek}, {Wittkowski},
      {Woillez}, {Wolff}, {Yazici}, {Ziegler}, \& {Zins}}]{gravity2017}
    {Gravity Collaboration}, {Abuter}, R., {Accardo}, M., {et~al.} 2017, \aap, 602,
      A94, \dodoi{10.1051/0004-6361/201730838}
    
    \bibitem[{{Gravity Collaboration} {et~al.}(2018){Gravity Collaboration},
      {Sturm}, {Dexter}, {Pfuhl}, {Stock}, {Davies}, {Lutz}, {Cl{\'e}net},
      {Eckart}, {Eisenhauer}, {Genzel}, {Gratadour}, {H{\"o}nig}, {Kishimoto},
      {Lacour}, {Millour}, {Netzer}, {Perrin}, {Peterson}, {Petrucci}, {Rouan},
      {Waisberg}, {Woillez}, {Amorim}, {Brandner}, {F{\"o}rster Schreiber},
      {Garcia}, {Gillessen}, {Ott}, {Paumard}, {Perraut}, {Scheithauer},
      {Straubmeier}, {Tacconi}, \& {Widmann}}]{gravity2018}
    {Gravity Collaboration}, {Sturm}, E., {Dexter}, J., {et~al.} 2018, \nat, 563,
      657, \dodoi{10.1038/s41586-018-0731-9}
    
    \bibitem[{{GRAVITY Collaboration} {et~al.}(2020){GRAVITY Collaboration},
      {Amorim}, {Brandner}, {Cl{\'e}net}, {Davies}, {de Zeeuw}, {Dexter}, {Eckart},
      {Eisenhauer}, {F{\"o}rster Schreiber}, {Gao}, {Garcia}, {Genzel},
      {Gillessen}, {Gratadour}, {H{\"o}nig}, {Kishimoto}, {Lacour}, {Lutz},
      {Millour}, {Netzer}, {Ott}, {Paumard}, {Perraut}, {Perrin}, {Peterson},
      {Petrucci}, {Pfuhl}, {Prieto}, {Rouan}, {Shangguan}, {Shimizu}, {Schartmann},
      {Sternberg}, {Straub}, {Straubmeier}, {Sturm}, {Tacconi}, {Tristram},
      {Vermot}, {von Fellenberg}, {Waisberg}, {Widmann}, \&
      {Woillez}}]{gravity2020}
    {GRAVITY Collaboration}, {Amorim}, A., {Brandner}, W., {et~al.} 2020, arXiv
      e-prints, arXiv:2009.08463.
    \newblock \doarXiv{2009.08463}
    
    \bibitem[{{Grier} {et~al.}(2013){Grier}, {Peterson}, {Horne}, {Bentz}, {Pogge},
      {Denney}, {De Rosa}, {Martini}, {Kochanek}, {Zu}, {Shappee}, {Siverd},
      {Beatty}, {Sergeev}, {Kaspi}, {Araya Salvo}, {Bird}, {Bord}, {Borman}, {Che},
      {Chen}, {Cohen}, {Dietrich}, {Doroshenko}, {Efimov}, {Free}, {Ginsburg},
      {Henderson}, {King}, {Mogren}, {Molina}, {Mosquera}, {Nazarov}, {Okhmat},
      {Pejcha}, {Rafter}, {Shields}, {Skowron}, {Szczygiel}, {Valluri}, \& {van
      Saders}}]{grier2013}
    {Grier}, C.~J., {Peterson}, B.~M., {Horne}, K., {et~al.} 2013, \apj, 764, 47,
      \dodoi{10.1088/0004-637X/764/1/47}
    
    \bibitem[{{Grier} {et~al.}(2017){Grier}, {Trump}, {Shen}, {Horne}, {Kinemuchi},
      {McGreer}, {Starkey}, {Brand t}, {Hall}, {Kochanek}, {Chen}, {Denney},
      {Greene}, {Ho}, {Homayouni}, {I-Hsiu Li}, {Pei}, {Peterson}, {Petitjean},
      {Schneider}, {Sun}, {AlSayyad}, {Bizyaev}, {Brinkmann}, {Brownstein},
      {Bundy}, {Dawson}, {Eftekharzadeh}, {Fernand ez-Trincado}, {Gao},
      {Hutchinson}, {Jia}, {Jiang}, {Oravetz}, {Pan}, {Paris}, {Ponder}, {Peters},
      {Rogerson}, {Simmons}, {Smith}, \& {Wang}}]{grier2017}
    {Grier}, C.~J., {Trump}, J.~R., {Shen}, Y., {et~al.} 2017, \apj, 851, 21,
      \dodoi{10.3847/1538-4357/aa98dc}
    
    \bibitem[{{Hopkins} {et~al.}(2007){Hopkins}, {Richards}, \&
      {Hernquist}}]{hopkins2007}
    {Hopkins}, P.~F., {Richards}, G.~T., \& {Hernquist}, L. 2007, \apj, 654, 731,
      \dodoi{10.1086/509629}
    
    \bibitem[{{Horne}(1994)}]{horne1994}
    {Horne}, K. 1994, in Astronomical Society of the Pacific Conference Series,
      Vol.~69, Reverberation Mapping of the Broad-Line Region in Active Galactic
      Nuclei, ed. P.~M. {Gondhalekar}, K.~{Horne}, \& B.~M. {Peterson}, 23
    
    \bibitem[{{Hu}(\noop{3001}in press)}]{hu2021}
    {Hu}, C. \noop{3001}in press, \apj
    
    \bibitem[{{Huang} \& {Wang}(2016)}]{huang2016}
    {Huang}, Q.-G., \& {Wang}, K. 2016, European Physical Journal C, 76, 506,
      \dodoi{10.1140/epjc/s10052-016-4352-x}
    
    \bibitem[{{Hubble}(1929)}]{hubble1929}
    {Hubble}, E. 1929, Proceedings of the National Academy of Science, 15, 168,
      \dodoi{10.1073/pnas.15.3.168}
    
    \bibitem[{{Kaspi} {et~al.}(2000){Kaspi}, {Smith}, {Netzer}, {Maoz}, {Jannuzi},
      \& {Giveon}}]{kaspi2000}
    {Kaspi}, S., {Smith}, P.~S., {Netzer}, H., {et~al.} 2000, \apj, 533, 631,
      \dodoi{10.1086/308704}
    
    \bibitem[{{Kelly} {et~al.}(2009){Kelly}, {Bechtold}, \&
      {Siemiginowska}}]{kelly2009}
    {Kelly}, B.~C., {Bechtold}, J., \& {Siemiginowska}, A. 2009, \apj, 698, 895,
      \dodoi{10.1088/0004-637X/698/1/895}
    
    \bibitem[{{Ko} \& {Tang}(2016)}]{ko2016}
    {Ko}, P., \& {Tang}, Y. 2016, Physics Letters B, 762, 462,
      \dodoi{10.1016/j.physletb.2016.10.001}
    
    \bibitem[{{Kreisch} {et~al.}(2020){Kreisch}, {Cyr-Racine}, \&
      {Dor{\'e}}}]{kreisch2020}
    {Kreisch}, C.~D., {Cyr-Racine}, F.-Y., \& {Dor{\'e}}, O. 2020, \prd, 101,
      123505, \dodoi{10.1103/PhysRevD.101.123505}
    
    \bibitem[{{Kun} {et~al.}(2015){Kun}, {Frey}, {Gab{\'a}nyi}, {Britzen}, {Cseh},
      \& {Gergely}}]{kun2015}
    {Kun}, E., {Frey}, S., {Gab{\'a}nyi}, K.~{\'E}., {et~al.} 2015, \mnras, 454,
      1290, \dodoi{10.1093/mnras/stv2049}
    
    \bibitem[{{Landt} {et~al.}(2008){Landt}, {Bentz}, {Ward}, {Elvis}, {Peterson},
      {Korista}, \& {Karovska}}]{landt2008}
    {Landt}, H., {Bentz}, M.~C., {Ward}, M.~J., {et~al.} 2008, \apjs, 174, 282,
      \dodoi{10.1086/522373}
    
    \bibitem[{{Li}(2020)}]{li2020}
    {Li}, Y.-R. 2020, {LiyrAstroph/CDNest: CDNest: A diffusive nested sampling code
      in C}, v0.2.0,  Zenodo, \dodoi{10.5281/zenodo.3884449}
    
    \bibitem[{{Li} {et~al.}(2013){Li}, {Wang}, {Ho}, {Du}, \& {Bai}}]{li2013}
    {Li}, Y.-R., {Wang}, J.-M., {Ho}, L.~C., {Du}, P., \& {Bai}, J.-M. 2013, \apj,
      779, 110, \dodoi{10.1088/0004-637X/779/2/110}
    
    \bibitem[{{Li} {et~al.}(2018){Li}, {Songsheng}, {Qiu}, {Hu}, {Du}, {Lu},
      {Huang}, {Bai}, {Bian}, {Yuan}, {Ho}, \& {Wang}}]{li2018}
    {Li}, Y.-R., {Songsheng}, Y.-Y., {Qiu}, J., {et~al.} 2018, \apj, 869, 137,
      \dodoi{10.3847/1538-4357/aaee6b}
    
    \bibitem[{{Liu} {et~al.}(2019){Liu}, {Liu}, {Dong}, {Zhou}, {Wang}, {Lu}, \&
      {Yuan}}]{liu2019}
    {Liu}, H.-Y., {Liu}, W.-J., {Dong}, X.-B., {et~al.} 2019, \apjs, 243, 21,
      \dodoi{10.3847/1538-4365/ab298b}
    
    \bibitem[{{Lu} {et~al.}(2016){Lu}, {Du}, {Hu}, {Li}, {Zhang}, {Wang}, {Huang},
      {Bi}, {Bai}, {Ho}, \& {Wang}}]{lu2016}
    {Lu}, K.-X., {Du}, P., {Hu}, C., {et~al.} 2016, \apj, 827, 118,
      \dodoi{10.3847/0004-637X/827/2/118}
    
    \bibitem[{{Lusso} {et~al.}(2012){Lusso}, {Comastri}, {Simmons}, {Mignoli},
      {Zamorani}, {Vignali}, {Brusa}, {Shankar}, {Lutz}, {Trump}, {Maiolino},
      {Gilli}, {Bolzonella}, {Puccetti}, {Salvato}, {Impey}, {Civano}, {Elvis},
      {Mainieri}, {Silverman}, {Koekemoer}, {Bongiorno}, {Merloni}, {Berta}, {Le
      Floc'h}, {Magnelli}, {Pozzi}, \& {Riguccini}}]{lusso2012}
    {Lusso}, E., {Comastri}, A., {Simmons}, B.~D., {et~al.} 2012, \mnras, 425, 623,
      \dodoi{10.1111/j.1365-2966.2012.21513.x}
    
    \bibitem[{{Lynden-Bell}(1969)}]{lyndenbell1969}
    {Lynden-Bell}, D. 1969, \nat, 223, 690, \dodoi{10.1038/223690a0}
    
    \bibitem[{{Marconi} {et~al.}(2003){Marconi}, {Maiolino}, \&
      {Petrov}}]{marconi2003}
    {Marconi}, A., {Maiolino}, R., \& {Petrov}, R.~G. 2003, \apss, 286, 245,
      \dodoi{10.1023/A:1026100831792}
    
    \bibitem[{{Pancoast} {et~al.}(2011){Pancoast}, {Brewer}, \&
      {Treu}}]{pancoast2011}
    {Pancoast}, A., {Brewer}, B.~J., \& {Treu}, T. 2011, \apj, 730, 139,
      \dodoi{10.1088/0004-637X/730/2/139}
    
    \bibitem[{{Pancoast} {et~al.}(2014{\natexlab{a}}){Pancoast}, {Brewer}, \&
      {Treu}}]{pancoast2014a}
    ---. 2014{\natexlab{a}}, \mnras, 445, 3055, \dodoi{10.1093/mnras/stu1809}
    
    \bibitem[{{Pancoast} {et~al.}(2014{\natexlab{b}}){Pancoast}, {Brewer}, {Treu},
      {Park}, {Barth}, {Bentz}, \& {Woo}}]{pancoast2014b}
    {Pancoast}, A., {Brewer}, B.~J., {Treu}, T., {et~al.} 2014{\natexlab{b}},
      \mnras, 445, 3073, \dodoi{10.1093/mnras/stu1419}
    
    \bibitem[{{Parkinson} {et~al.}(2011){Parkinson}, {Mukherjee}, \&
      {Liddle}}]{parkinson2011}
    {Parkinson}, D., {Mukherjee}, P., \& {Liddle}, A. 2011, {CosmoNest:
      Cosmological Nested Sampling}.
    \newblock \doeprint{1110.019}
    
    \bibitem[{{Perlmutter} {et~al.}(1999){Perlmutter}, {Aldering}, {Goldhaber},
      {Knop}, {Nugent}, {Castro}, {Deustua}, {Fabbro}, {Goobar}, {Groom}, {Hook},
      {Kim}, {Kim}, {Lee}, {Nunes}, {Pain}, {Pennypacker}, {Quimby}, {Lidman},
      {Ellis}, {Irwin}, {McMahon}, {Ruiz-Lapuente}, {Walton}, {Schaefer}, {Boyle},
      {Filippenko}, {Matheson}, {Fruchter}, {Panagia}, {Newberg}, {Couch}, \&
      {Project}}]{perlmutter1999}
    {Perlmutter}, S., {Aldering}, G., {Goldhaber}, G., {et~al.} 1999, \apj, 517,
      565, \dodoi{10.1086/307221}
    
    \bibitem[{{Peterson}(1993)}]{peterson1993}
    {Peterson}, B.~M. 1993, \pasp, 105, 247, \dodoi{10.1086/133140}
    
    \bibitem[{{Peterson} {et~al.}(1998){Peterson}, {Wanders}, {Bertram}, {Hunley},
      {Pogge}, \& {Wagner}}]{peterson1998}
    {Peterson}, B.~M., {Wanders}, I., {Bertram}, R., {et~al.} 1998, \apj, 501, 82,
      \dodoi{10.1086/305813}
    
    \bibitem[{{Peterson} {et~al.}(2004){Peterson}, {Ferrarese}, {Gilbert}, {Kaspi},
      {Malkan}, {Maoz}, {Merritt}, {Netzer}, {Onken}, {Pogge}, {Vestergaard}, \&
      {Wandel}}]{peterson2004}
    {Peterson}, B.~M., {Ferrarese}, L., {Gilbert}, K.~M., {et~al.} 2004, \apj, 613,
      682, \dodoi{10.1086/423269}
    
    \bibitem[{{Petrov}(1989)}]{petrov1989}
    {Petrov}, R.~G. 1989, in NATO Advanced Science Institutes (ASI) Series C, Vol.
      274, NATO Advanced Science Institutes (ASI) Series C, ed. D.~M. {Alloin} \&
      J.~M. {Mariotti}, 249
    
    \bibitem[{{Petrov} {et~al.}(2001){Petrov}, {Malbet}, {Richichi}, {Hofmann},
      {Mourard}, \& {Amber Consortium}}]{petrov2001}
    {Petrov}, R.~G., {Malbet}, F., {Richichi}, A., {et~al.} 2001, Comptes Rendus
      Physique, 2, 67.
    \newblock \doarXiv{astro-ph/0507398}
    
    \bibitem[{{Planck Collaboration} {et~al.}(2018){Planck Collaboration},
      {Aghanim}, {Akrami}, {Ashdown}, {Aumont}, {Baccigalupi}, {Ballardini},
      {Banday}, {Barreiro}, {Bartolo}, {Basak}, {Battye}, {Benabed}, {Bernard},
      {Bersanelli}, {Bielewicz}, {Bock}, {Bond}, {Borrill}, {Bouchet}, {Boulanger},
      {Bucher}, {Burigana}, {Butler}, {Calabrese}, {Cardoso}, {Carron},
      {Challinor}, {Chiang}, {Chluba}, {Colombo}, {Combet}, {Contreras}, {Crill},
      {Cuttaia}, {de Bernardis}, {de Zotti}, {Delabrouille}, {Delouis}, {Di
      Valentino}, {Diego}, {Dor{\'e}}, {Douspis}, {Ducout}, {Dupac}, {Dusini},
      {Efstathiou}, {Elsner}, {En{\ss}lin}, {Eriksen}, {Fantaye}, {Farhang},
      {Fergusson}, {Fernandez-Cobos}, {Finelli}, {Forastieri}, {Frailis},
      {Fraisse}, {Franceschi}, {Frolov}, {Galeotta}, {Galli}, {Ganga},
      {G{\'e}nova-Santos}, {Gerbino}, {Ghosh}, {Gonz{\'a}lez-Nuevo}, {G{\'o}rski},
      {Gratton}, {Gruppuso}, {Gudmundsson}, {Hamann}, {Handley}, {Hansen},
      {Herranz}, {Hildebrandt}, {Hivon}, {Huang}, {Jaffe}, {Jones}, {Karakci},
      {Keih{\"a}nen}, {Keskitalo}, {Kiiveri}, {Kim}, {Kisner}, {Knox},
      {Krachmalnicoff}, {Kunz}, {Kurki-Suonio}, {Lagache}, {Lamarre}, {Lasenby},
      {Lattanzi}, {Lawrence}, {Le Jeune}, {Lemos}, {Lesgourgues}, {Levrier},
      {Lewis}, {Liguori}, {Lilje}, {Lilley}, {Lindholm}, {L{\'o}pez-Caniego},
      {Lubin}, {Ma}, {Mac{\'\i}as-P{\'e}rez}, {Maggio}, {Maino}, {Mandolesi},
      {Mangilli}, {Marcos-Caballero}, {Maris}, {Martin}, {Martinelli},
      {Mart{\'\i}nez-Gonz{\'a}lez}, {Matarrese}, {Mauri}, {McEwen}, {Meinhold},
      {Melchiorri}, {Mennella}, {Migliaccio}, {Millea}, {Mitra},
      {Miville-Desch{\^e}nes}, {Molinari}, {Montier}, {Morgante}, {Moss}, {Natoli},
      {N{\o}rgaard-Nielsen}, {Pagano}, {Paoletti}, {Partridge}, {Patanchon},
      {Peiris}, {Perrotta}, {Pettorino}, {Piacentini}, {Polastri}, {Polenta},
      {Puget}, {Rachen}, {Reinecke}, {Remazeilles}, {Renzi}, {Rocha}, {Rosset},
      {Roudier}, {Rubi{\~n}o-Mart{\'\i}n}, {Ruiz-Granados}, {Salvati}, {Sandri},
      {Savelainen}, {Scott}, {Shellard}, {Sirignano}, {Sirri}, {Spencer},
      {Sunyaev}, {Suur-Uski}, {Tauber}, {Tavagnacco}, {Tenti}, {Toffolatti},
      {Tomasi}, {Trombetti}, {Valenziano}, {Valiviita}, {Van Tent}, {Vibert},
      {Vielva}, {Villa}, {Vittorio}, {Wand elt}, {Wehus}, {White}, {White},
      {Zacchei}, \& {Zonca}}]{planck2018}
    {Planck Collaboration}, {Aghanim}, N., {Akrami}, Y., {et~al.} 2018, arXiv
      e-prints, arXiv:1807.06209.
    \newblock \doarXiv{1807.06209}
    
    \bibitem[{{Raimundo} {et~al.}(2019){Raimundo}, {Pancoast}, {Vestergaard},
      {Goad}, \& {Barth}}]{raimundo2019}
    {Raimundo}, S.~I., {Pancoast}, A., {Vestergaard}, M., {Goad}, M.~R., \&
      {Barth}, A.~J. 2019, \mnras, 489, 1899, \dodoi{10.1093/mnras/stz2243}
    
    \bibitem[{{Raimundo} {et~al.}(2020){Raimundo}, {Vestergaard}, {Goad}, {Grier},
      {Williams}, {Peterson}, \& {Treu}}]{raimundo2020}
    {Raimundo}, S.~I., {Vestergaard}, M., {Goad}, M.~R., {et~al.} 2020, \mnras,
      493, 1227, \dodoi{10.1093/mnras/staa285}
    
    \bibitem[{{Rakshit} {et~al.}(2015){Rakshit}, {Petrov}, {Meilland}, \&
      {H{\"o}nig}}]{rakshit2015}
    {Rakshit}, S., {Petrov}, R.~G., {Meilland}, A., \& {H{\"o}nig}, S.~F. 2015,
      \mnras, 447, 2420, \dodoi{10.1093/mnras/stu2613}
    
    \bibitem[{{Rameez} \& {Sarkar}(2019)}]{rameez2019}
    {Rameez}, M., \& {Sarkar}, S. 2019, arXiv e-prints, arXiv:1911.06456.
    \newblock \doarXiv{1911.06456}
    
    \bibitem[{{Rees}(1984)}]{rees1984}
    {Rees}, M.~J. 1984, \araa, 22, 471, \dodoi{10.1146/annurev.aa.22.090184.002351}
    
    \bibitem[{{Refsdal}(1964)}]{refsdal1964}
    {Refsdal}, S. 1964, \mnras, 128, 307, \dodoi{10.1093/mnras/128.4.307}
    
    \bibitem[{{Riess} {et~al.}(2019){Riess}, {Casertano}, {Yuan}, {Macri}, \&
      {Scolnic}}]{riess2019}
    {Riess}, A.~G., {Casertano}, S., {Yuan}, W., {Macri}, L.~M., \& {Scolnic}, D.
      2019, \apj, 876, 85, \dodoi{10.3847/1538-4357/ab1422}
    
    \bibitem[{{Riess} {et~al.}(1998){Riess}, {Filippenko}, {Challis},
      {Clocchiatti}, {Diercks}, {Garnavich}, {Gilliland}, {Hogan}, {Jha},
      {Kirshner}, {Leibundgut}, {Phillips}, {Reiss}, {Schmidt}, {Schommer},
      {Smith}, {Spyromilio}, {Stubbs}, {Suntzeff}, \& {Tonry}}]{riess1998}
    {Riess}, A.~G., {Filippenko}, A.~V., {Challis}, P., {et~al.} 1998, \aj, 116,
      1009, \dodoi{10.1086/300499}
    
    \bibitem[{{Rose} {et~al.}(2020){Rose}, {Rubin}, {Strolger}, \&
      {Garnavich}}]{rose2020}
    {Rose}, B.~M., {Rubin}, D., {Strolger}, L., \& {Garnavich}, P. 2020, in
      American Astronomical Society Meeting Abstracts, American Astronomical
      Society Meeting Abstracts, 108.01
    
    \bibitem[{{Schutz}(1986)}]{schutz1986}
    {Schutz}, B.~F. 1986, \nat, 323, 310, \dodoi{10.1038/323310a0}
    
    \bibitem[{{Shaw} {et~al.}(2007){Shaw}, {Bridges}, \& {Hobson}}]{shaw2007}
    {Shaw}, J.~R., {Bridges}, M., \& {Hobson}, M.~P. 2007, \mnras, 378, 1365,
      \dodoi{10.1111/j.1365-2966.2007.11871.x}
    
    \bibitem[{{Shen} {et~al.}(2016){Shen}, {Horne}, {Grier}, {Peterson}, {Denney},
      {Trump}, {Sun}, {Brandt}, {Kochanek}, {Dawson}, {Green}, {Greene}, {Hall},
      {Ho}, {Jiang}, {Kinemuchi}, {McGreer}, {Petitjean}, {Richards}, {Schneider},
      {Strauss}, {Tao}, {Wood-Vasey}, {Zu}, {Pan}, {Bizyaev}, {Ge}, {Oravetz}, \&
      {Simmons}}]{shen2016}
    {Shen}, Y., {Horne}, K., {Grier}, C.~J., {et~al.} 2016, \apj, 818, 30,
      \dodoi{10.3847/0004-637X/818/1/30}
    
    \bibitem[{{Skilling}(2004)}]{skilling2004}
    {Skilling}, J. 2004, in American Institute of Physics Conference Series, Vol.
      735, Bayesian Inference and Maximum Entropy Methods in Science and
      Engineering: 24th International Workshop on Bayesian Inference and Maximum
      Entropy Methods in Science and Engineering, ed. R.~{Fischer}, R.~{Preuss}, \&
      U.~V. {Toussaint}, 395--405
    
    \bibitem[{{Skilling}(2006)}]{john2006}
    {Skilling}, J. 2006, Bayesian Analysis, 1, 833, \dodoi{10.1214/06-BA127}
    
    \bibitem[{{Soares-Santos} {et~al.}(2017){Soares-Santos}, {Holz}, {Annis},
      {Chornock}, {Herner}, {Berger}, {Brout}, {Chen}, {Kessler}, {Sako}, {Allam},
      {Tucker}, {Butler}, {Palmese}, {Doctor}, {Diehl}, {Frieman}, {Yanny}, {Lin},
      {Scolnic}, {Cowperthwaite}, \& {Neilsen}}]{soaressantos2017}
    {Soares-Santos}, M., {Holz}, D.~E., {Annis}, J., {et~al.} 2017, \apjl, 848,
      L16, \dodoi{10.3847/2041-8213/aa9059}
    
    \bibitem[{{Songsheng} {et~al.}(2019{\natexlab{a}}){Songsheng}, {Wang}, \&
      {Li}}]{songsheng2019b}
    {Songsheng}, Y.-Y., {Wang}, J.-M., \& {Li}, Y.-R. 2019{\natexlab{a}}, \apj,
      883, 184, \dodoi{10.3847/1538-4357/ab3c5e}
    
    \bibitem[{{Songsheng} {et~al.}(2019{\natexlab{b}}){Songsheng}, {Wang}, {Li}, \&
      {Du}}]{songsheng2019}
    {Songsheng}, Y.-Y., {Wang}, J.-M., {Li}, Y.-R., \& {Du}, P. 2019{\natexlab{b}},
      \apj, 881, 140, \dodoi{10.3847/1538-4357/ab2e00}
    
    \bibitem[{{Suyu} {et~al.}(2017){Suyu}, {Bonvin}, {Courbin}, {Fassnacht},
      {Rusu}, {Sluse}, {Treu}, {Wong}, {Auger}, {Ding}, {Hilbert}, {Marshall},
      {Rumbaugh}, {Sonnenfeld}, {Tewes}, {Tihhonova}, {Agnello}, {Blandford},
      {Chen}, {Collett}, {Koopmans}, {Liao}, {Meylan}, \& {Spiniello}}]{suyu2017}
    {Suyu}, S.~H., {Bonvin}, V., {Courbin}, F., {et~al.} 2017, \mnras, 468, 2590,
      \dodoi{10.1093/mnras/stx483}
    
    \bibitem[{{Verde} {et~al.}(2019){Verde}, {Treu}, \& {Riess}}]{verde2019}
    {Verde}, L., {Treu}, T., \& {Riess}, A.~G. 2019, Nature Astronomy, 3, 891,
      \dodoi{10.1038/s41550-019-0902-0}
    
    \bibitem[{{Wang} {et~al.}(2020){Wang}, {Songsheng}, {Li}, {Du}, \&
      {Zhang}}]{wang2020}
    {Wang}, J.-M., {Songsheng}, Y.-Y., {Li}, Y.-R., {Du}, P., \& {Zhang}, Z.-X.
      2020, Nature Astronomy, 4, 517, \dodoi{10.1038/s41550-019-0979-5}
    
    \bibitem[{{Williams} {et~al.}(2018){Williams}, {Pancoast}, {Treu}, {Brewer},
      {Barth}, {Bennert}, {Buehler}, {Canalizo}, {Cenko}, {Clubb}, {Cooper},
      {Filippenko}, {Gates}, {Hoenig}, {Joner}, {Kandrashoff}, {Laney}, {Lazarova},
      {Li}, {Malkan}, {Rex}, {Silverman}, {Tollerud}, {Walsh}, \&
      {Woo}}]{williams2018}
    {Williams}, P.~R., {Pancoast}, A., {Treu}, T., {et~al.} 2018, \apj, 866, 75,
      \dodoi{10.3847/1538-4357/aae086}
    
    \bibitem[{{Wong} {et~al.}(2020){Wong}, {Suyu}, {Chen}, {Rusu}, {Millon},
      {Sluse}, {Bonvin}, {Fassnacht}, {Taubenberger}, {Auger}, {Birrer}, {Chan},
      {Courbin}, {Hilbert}, {Tihhonova}, {Treu}, {Agnello}, {Ding}, {Jee},
      {Komatsu}, {Shajib}, {Sonnenfeld}, {Bland ford}, {Koopmans}, {Marshall}, \&
      {Meylan}}]{wong2020}
    {Wong}, K.~C., {Suyu}, S.~H., {Chen}, G. C.~F., {et~al.} 2020, \mnras,
      \dodoi{10.1093/mnras/stz3094}
    
    \bibitem[{{Xiao} {et~al.}(2018){Xiao}, {Du}, {Horne}, {Hu}, {Li}, {Huang},
      {Lu}, {Qiu}, {Wang}, {Bai}, {Bian}, {Ho}, {Yuan}, {Wang}, \& {SEAMBH
      Collaboration}}]{xiao2018}
    {Xiao}, M., {Du}, P., {Horne}, K., {et~al.} 2018, \apj, 864, 109,
      \dodoi{10.3847/1538-4357/aad5e1}
    
    \bibitem[{{Zhang} {et~al.}(2019){Zhang}, {Du}, {Smith}, {Zhao}, {Hu}, {Xiao},
      {Li}, {Huang}, {Wang}, {Bai}, {Ho}, \& {Wang}}]{zhang2019}
    {Zhang}, Z.-X., {Du}, P., {Smith}, P.~S., {et~al.} 2019, \apj, 876, 49,
      \dodoi{10.3847/1538-4357/ab1099}
    
    \bibitem[{{Zhao} {et~al.}(2017){Zhao}, {Raveri}, {Pogosian}, {Wang},
      {Crittenden}, {Handley}, {Percival}, {Beutler}, {Brinkmann}, {Chuang},
      {Cuesta}, {Eisenstein}, {Kitaura}, {Koyama}, {L'Huillier}, {Nichol}, {Pieri},
      {Rodriguez-Torres}, {Ross}, {Rossi}, {S{\'a}nchez}, {Shafieloo}, {Tinker},
      {Tojeiro}, {Vazquez}, \& {Zhang}}]{zhao2017}
    {Zhao}, G.-B., {Raveri}, M., {Pogosian}, L., {et~al.} 2017, Nature Astronomy,
      1, 627, \dodoi{10.1038/s41550-017-0216-z}
    
    \bibitem[{{Zu} {et~al.}(2013){Zu}, {Kochanek}, {Koz{\l}owski}, \&
      {Udalski}}]{zu2013}
    {Zu}, Y., {Kochanek}, C.~S., {Koz{\l}owski}, S., \& {Udalski}, A. 2013, \apj,
      765, 106, \dodoi{10.1088/0004-637X/765/2/106}
    
\end{thebibliography}

\appendix
\section{Mock data}\label{sec:mock data}
\begin{figure}
  \plotone{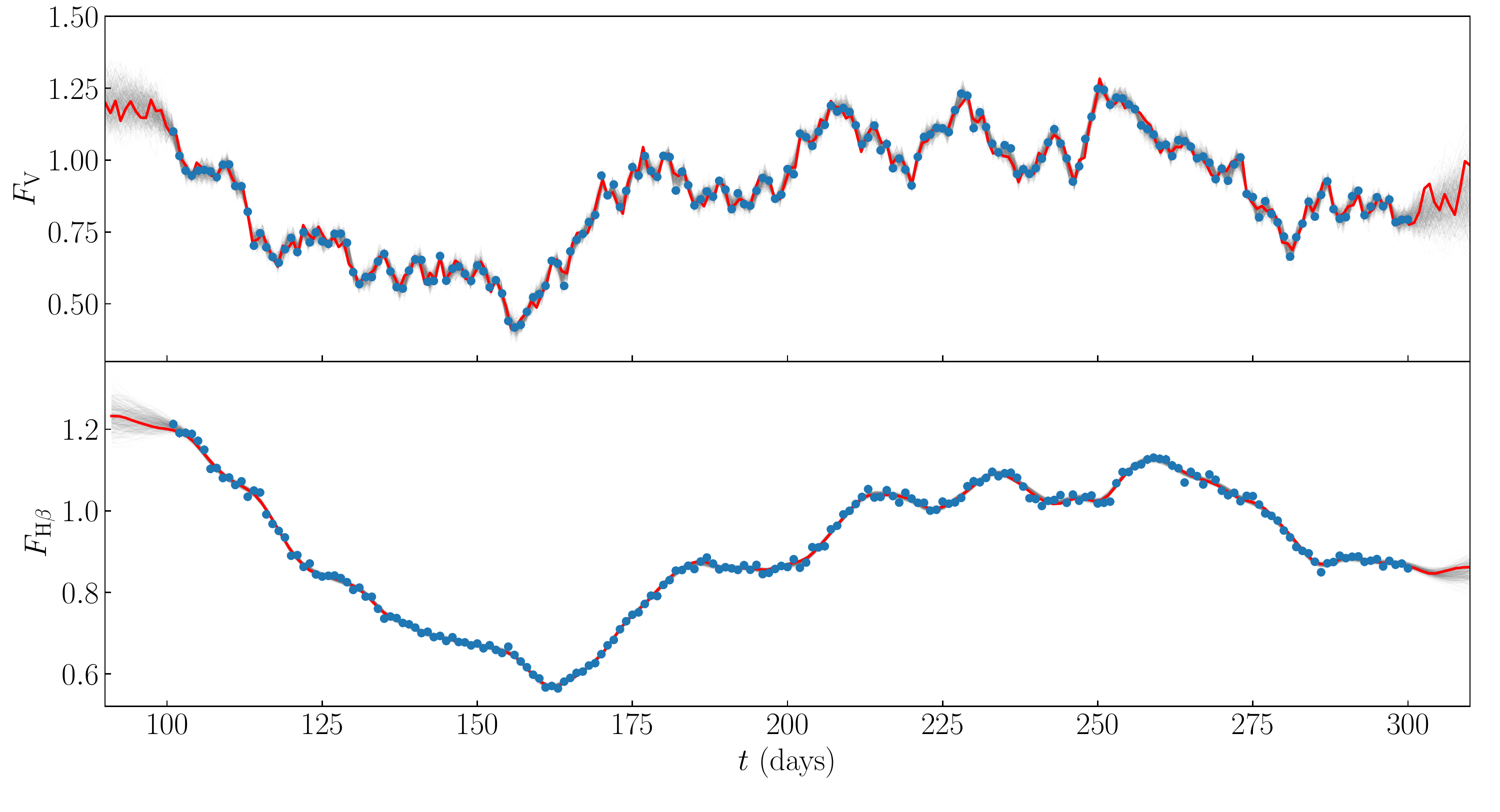}
  \caption{An example of mock light curves we generated and its fitting to 
  the BLR model. The first row is continuum variation and the second is 
  emission line variation. Error bars for data points reflect 1 $\sigma$ 
  uncertainties. The thick red lines are the best fitting, while the thin gray \
  lines are fittings using model parameters drawn from the probability 
  distribution of model parameters.
  \label{fig:lc}}
\end{figure}

Here, we present an example of mock data generation and its fitting for 
illustration. We use the parameterized BLR model describe in subsection \ref
{sec:model}, and values of its parameters are taken to be the fiducial values 
listed in Table \ref{tab:BLR}. The continuum variation is generated by the DRW 
model with timescale $\tau_{\rm d} = 60 {\rm\,d}$ and amplitude $\sigma_{\rm d} 
= 0.25$. Then it is convolved with the velocity-resolved transfer 
function determined by BLR model to obtain the light curve of emission line. 
Both light curves are sampled with $1$ day cadence and last for $200$ days. The 
relative uncertainties of each data point are $0.5\%$ for continuum and $1\%$ 
for emission line. Blue points with errorbars in Fig \ref{fig:lc} are mock 
light curves we generated.

\begin{figure}
    \plotone{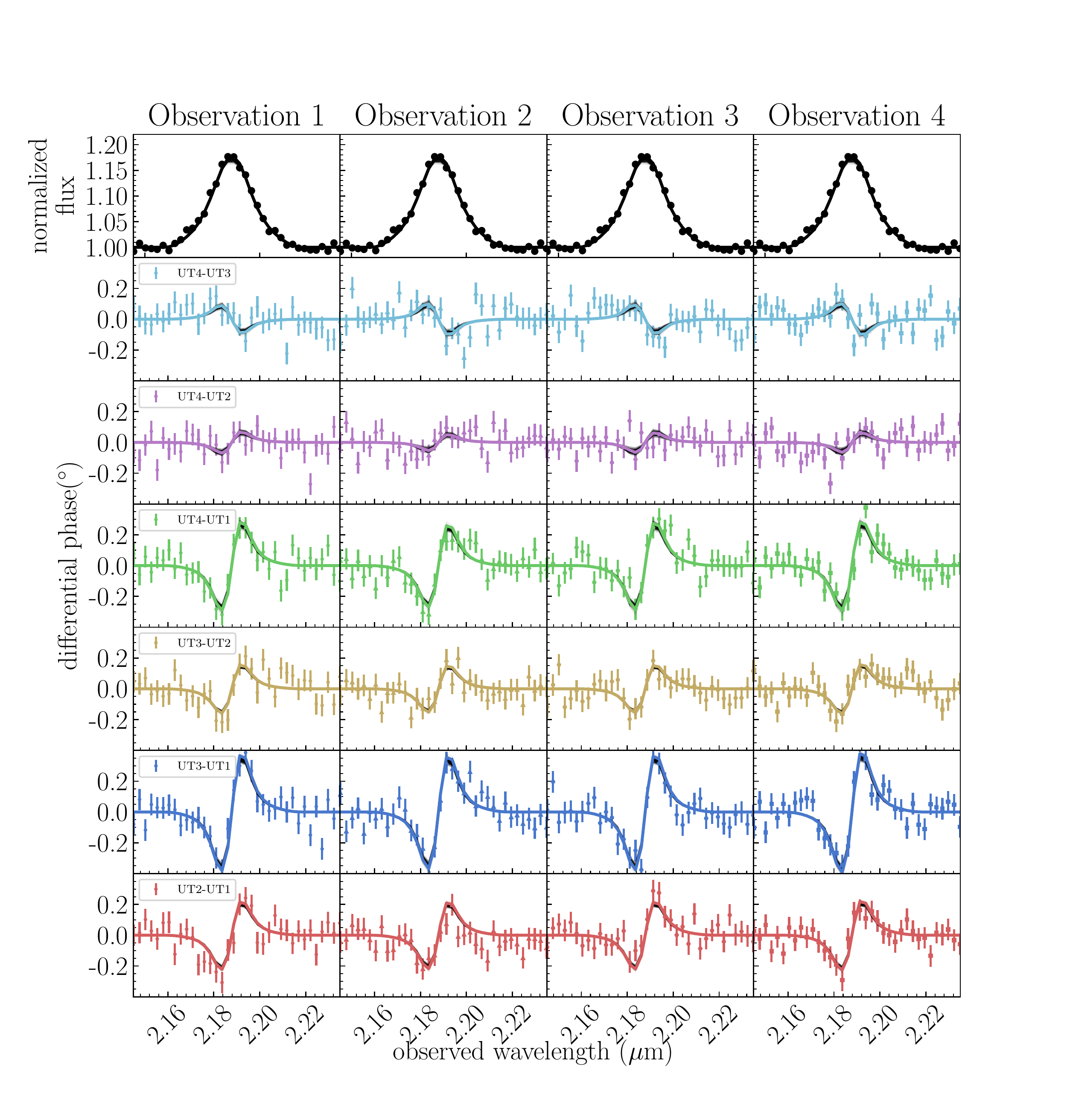}
    \caption{An example of mock profile and DPC we generated and its fitting to 
    the BLR model. The first row is the Br$\gamma$ profile, and the second to 
    the last rows are corresponding DPCs for each baseline. The object is 
    observed for four times. Error bars for data points reflect 1 $\sigma$ 
    uncertainties. The thick solid lines are the best fitting while the thin 
    gray lines are fittings using model parameters drawn from the probability 
    distribution of model parameters.
    \label{fig:dpc}}
\end{figure}

The profile and DPCs are simulated using the same BLR model. The emission 
line used for SA is Br$\gamma$. To normalize the profile, we assume the 
equivalent width of the line is $40\text{\AA}$ in rest frame. The redshift of 
the object is $0.01$, shifting the central wavelength of the Br$\gamma$ 
to $2.18766{\rm\,\mu m}$. There are $40$ wavelength bins between $2.14$ and $2.
24 {\rm\,\mu m}$, and the profile and DPC are broadened by a Gaussian with $
{\rm FWHM} = 4{\rm\,nm}$. The relative uncertainty of the profile and DPC are 
$0.5\%$ and $20\%$ respectively. The object is observed for four times with six 
baselines, and the configuration of baselines is the same as that of Extended 
Data Fig. 1 (b) in \citet{gravity2018}. Points with errorbars in Fig \ref
{fig:dpc} are mock profile and differential phase curves we generated.

Bayesian analysis is applied to fit the combined mock data jointly to the BLR 
model and obtain the posterior probability distribution of model parameters, as 
described in subsection \ref{sec:joint analysis}. DNest algorithm is used to 
sample the posterior distribution \ref{eq:posterior} \citep{brewer2018,li2020}, 
generating a posterior sample of model parameters and associated light curves, 
profile and DPCs. The reconstructed curves with smallest $\chi^2$ (the best 
fitting) are thick solid lines shown in Fig \ref{fig:lc} and \ref{fig:dpc}, 
while those randomly drawn from the posterior sample are thin gray lines.

\section{Mock sample}
\label{sec:mock sample}
\begin{figure}
    \plotone{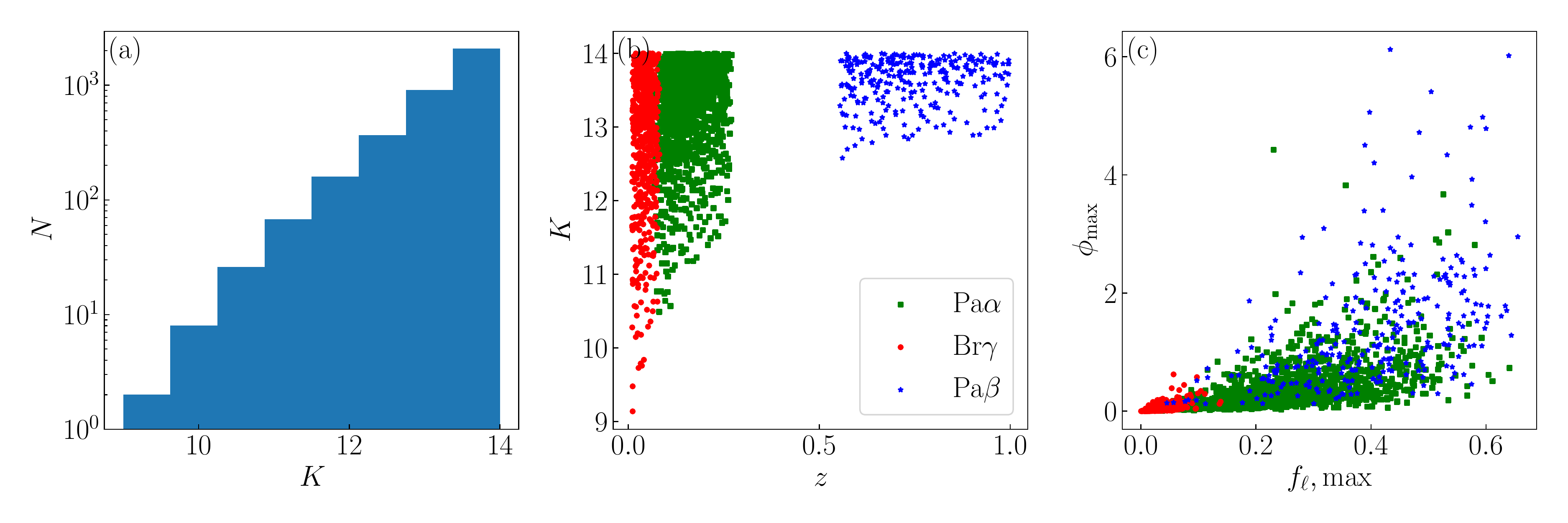}
    \caption{(a) The number of Type I AGNs with appropriate declinations for 
    VLTI in each bin of $K$ band magnitudes.
    (b) The distribution of redshifts and $K$ band magnitude of objects with $
    {\rm Br}\gamma$, ${\rm Pa}\alpha$ or ${\rm Pa}\beta$ emission lines within 
    the spectral coverage of GRAVITY. 
    (c) The distribution of line intensities $f_{\ell{\rm,max}}$ and phase 
    signal amplitudes $\phi_{\rm max}$ of objects in (b).
    \label{fig:stats}}
\end{figure}

In order to study the constraint ability on $H_0$ by joint analysis of SARM 
data from a large sample of type I AGNs, we need to generate a mock sample 
using realistic luminosity functions of quasars. We first divide the redshift 
(up to $z=1$) and bolometric luminosity (from $10^{41.6}$ to $10^{51.5} {\rm 
erg\,s^{-1}}$) into short bins. Then we estimate the number of quasars in each 
bin using the redshift-dependent bolometric quasar luminosity function\footnote
{A quasar luminosity function calculator script is available at \url{http://www.
tapir.caltech.edu/~phopkins/Site/qlf.html}} in \citet{hopkins2007}. Assuming 
the inclination of AGNs are isotropic and broad emission line can be observed 
when $i < 45\degr$, only $29.3\%$ of them are type I. The distribution of AGNs 
on celestial sphere is also isotropic. When the difference between the 
declination of the target and latitude of VLTI is smaller than $45\degr$, the 
object can be observed by GRAVITY. Thus, only $64.3\%$ of the remaining type I 
AGNs are selected. For each object in our mock list, we calculate the $L_{5100}
$ luminosity and $K$ band magnitude using the bolometric luminosity dependent 
spectral energy distribution provided by \citet{hopkins2007}. Objects whose 
redshifts are smaller than $0.01$ are discarded since their peculiar motions 
may be comparable to the Hubble flow. We also remove those with bolometric 
luminosities larger than $10^{48} {\rm erg\,s^{-1}}$ because their variations 
are too slow for a typical RM campaign and those with $K$ band magnitude larger 
than $13$ as they are too faint. The distribution of $K$ band magnitudes of our 
mock sample is shown in Fig. \ref{fig:stats}(a).

Before simulating line profiles and DPCs for objects in our mock sample, we 
need to know values of model parameters listed in Table \ref{tab:BLR} of each 
BLR. The standard $R-L$ relation in \citet{bentz2013} is applied to estimate 
the size of BLR, while luminosity dependent Eddington ratio of Type I AGNs 
measured by \citet{lusso2012} are used to infer the black hole mass. Note that 
dispersions in those relations are also included. Other model parameters are 
assigned randomly according to the the lase columns of Table \ref{tab:BLR}.

The spectral coverage of GRAVITY is $1.98-2.40 {\rm\,\mu m}$. In hydrogen 
emission lines, only Br$\gamma$ ($2.166 {\rm\,\mu m}$), Pa$\alpha$ 
($1.875{\rm \,\mu m}$) and Pa$\beta$ ($1.282{\rm \,\mu m}$) are possible 
to be observed by GRAVITY for objects with $z < 1$. \cite{landt2008} studied  
near-infrared broad emission line properties of 23 well-known Type I AGNs. We 
assume distributions of equivalent widths of ${\rm Br}\gamma$, ${\rm Pa}\alpha$ 
and ${\rm Pa}\beta$ in our mock samples are Gaussian with means and standard 
deviations determined by samples in \cite{landt2008}. For each object we 
calculate line profiles in observer's frame of those three emission lines. We 
adopt the emission line with largest equivalent width within the spectral 
coverage of GRAVITY as target line. Objects without appropriate emission lines 
are discarded. The distribution of redshifts and $K$ band magnitude of objects 
with emission lines within the spectral coverage is shown in Fig. \ref
{fig:stats}(b). When $z \lesssim 0.1$, ${\rm Br}\gamma$ can be observed by 
GRAVITY, and $K$ band magnitudes can be as bright as $9$. When $0.1 \lesssim z 
\lesssim 0.3$, ${\rm Pa}\alpha$ is chosen, and the $K$ band magnitude of the 
brightest one is about $10.5$. When $0.6 \lesssim z \lesssim 0.8$, ${\rm Pa}
\beta$ is , and $K$ band magnitudes are around $14$. Note that objects are 
removed if FWHMs of their profiles are less than $1500{\rm\,km\,s^{-1}}$ due to 
the limited spectral resolution of GRAVITY ($\sim 500$ when observing AGNs) or 
if inclinations are less than $10\degr$ because of high uncertainties in 
distance measurements of face-on objects.

Finally, to determine the projected lengths of baselines and calculate the 
DPC for each object, we assume all of them are observed when reaching their 
zeniths for simplicity. The distribution of line intensities $f_{\ell{\rm,max}}
$ and phase signal amplitudes $\phi_{\rm max}$ is shown in Fig. \ref{fig:stats}
(c). For ${\rm Br}\gamma$ lines, $f_{\ell{\rm,max}} $ is mostly less than $0.1$ 
and $\phi_{\rm max}$ is less than $0.2 \degr$, which is compatible to the 
latest observation of IRAS 09149-6206 by GRAVITY \citep{gravity2020}. For ${\rm 
Pa}\alpha$ lines, $f_{\ell{\rm,max}}$ is about $0.2-0.4$ and $\phi_{\rm max}$ 
is about $0.2-0.6 \degr$, which is compatible to the recent observation of 3C 
273 by GRAVITY \citep{gravity2018}.

{\cblue
\section{Diffusive nested sampling}
\label{sec:DNest}
Nested sampling was proposed by \cite{skilling2004} to evaluate the evidence 
$Z$ of a model $M$:
\begin{equation}
    Z \equiv P(\mathscr{D}|M) = \int P(\mathscr{D} | \bm{\Theta},M) P(\bm{\Theta}|M) \dd{\bm{\Theta}},
\end{equation}
where $\mathscr{D}$ and $\bm{\Theta}$ represent the data set and model 
parameters respectively, $P(\bm{\Theta}|M)$ is the prior probability 
distribution of parameters $\bm{\Theta}$ in model $M$ and $P(\mathscr{D} | \bm
{\Theta},M)$ is the likelihood function.

Nested sampling starts with $n$ points $\bm{\Theta}_i$ sampled from prior $P(\bm
{\Theta}|M)$. The likelihood of each point $L(\bm{\Theta}_i) \equiv P(\mathscr
{D} | \bm{\Theta}_i,M)$ is evaluated. The minimum of likelihoods $L_1$ and 
corresponding particle is saved. Then this particle will be replaced by a 
new one drawn from the prior probability distribution but under a constraint $L
(\bm{\Theta}) > L_1$ via Markov chain Monte Carlo method. Again, the minimum of 
likelihoods of the living $n$ particles $L_2$ is saved and iteration continues. 
As nested likelihood levels $L_1 < L_2 < \cdots$ are created, particles move 
progressively towards higher likelihoods. The posterior distribution of $\bm
{\Theta}$ can be obtained as a byproduct by recording positions and likelihoods 
of particles in the process.

There are several variants of nested samling, such as \texttt{CosmoNest}\citep
{parkinson2011} and \texttt{MultiNest}\citep{feroz2009}. Diffusive nested 
sampling \citep{brewer2011} is the one that makes improvements to 
the original method to overcome its drawbacks when sampling multimodal or 
highly correlated distributions. In the process of creating levels, particles 
at high levels can diffuse to lower levels. If a distribution has isolate 
islands with high likelihoods, particles in classic nested sampling may be 
stuck in one island and fail to explore other islands. However, particles in 
diffusive nested sampling can diffuse to lower levels where the distribution is 
sufficently broad without isolate islands so that particles can easily move 
over the whole parameter space.

Diffusive nested sampling peforms well to overcome high dimensions, multimodal 
distributions and phase changes. It has been successfully applied to fit the 
broad line region model to reverberation mapping and spectroastrometry data to 
measure the broad line region size, black hole mass, and distance to the quasar 
\citep[e.g.][]{pancoast2014b, li2018, wang2020}.
}
\end{document}